# A Comparative Survey of Optical Wireless Technologies: Architectures and Applications

Mostafa Zaman Chowdhury, Md. Tanvir Hossan, Amirul Islam, and Yeong Min Jang

*Abstract*— New high-data-rate multimedia services and applications are evolving continuously and exponentially increasing the demand for wireless capacity of fifth-generation (5G) and beyond. The existing radio frequency (RF) communication spectrum is insufficient to meet the demands of future high-data-rate 5G services. Optical wireless communication (OWC), which uses an ultra-wide range of unregulated spectrum, has emerged as a promising solution to overcome the RF spectrum crisis. It has attracted growing research interest worldwide in the last decade for indoor and outdoor applications. OWC offloads huge data traffic applications from RF networks. A 100 Gbps data rate has already been demonstrated through OWC. It offers services indoors as well as outdoors, and communication distances range from several nm to more than 10,000 km. This paper provides a technology overview and a review on optical wireless technologies such as visible light communication, light fidelity, optical camera communication, free space optical communication, and light detection and ranging. We survey the key technologies for understanding OWC and present state-of-the-art criteria in aspects such as classification, spectrum use, architecture, and applications. The key contribution of this paper is to clarify the differences among different promising optical wireless technologies and between these technologies and their corresponding similar existing RF technologies.

## I. INTRODUCTION

Fifth-generation (5G) communication is the next step in mobile telecommunication standards. It will offer new services with ultra-high system capacity, massive device connectivity, ultra-low latency, ultra-high security, ultra-low energy consumption, and extremely high quality of experience [1]–[6]. It is expected that 5G communication will comprise ultra-dense heterogeneous networks for which mobile data volume per area will be 1000 times and the number of connected wireless devices will be 100 times higher compared to those in existing wireless networks [1]. Therefore, upcoming networks must Mostafa Zaman Chowdhury is with Department of Electronics Engineering, Kookmin University, Korea (e-mail: mzaman@kookmin.ac.kr).
Md. Tanvir Hossan is with Department of Electronics Engineering, Kookmin University, Korea (e-mail: tanvir.hossan.shaon@gmail.com).
Amirul Islam is with Department of Electronics Engineering, Kookmin University, Korea (e-mail: amirul.0903118@gmail.com).
Yeong Min Jang is with Department of Electronics Engineering, Kookmin University, Korea (e-mail: yjang@kookmin.ac.kr).

have the capability to support high user data rates, low power consumption, and negligible end-to-end delays [1]–[6]. High-capacity backhaul connectivity is essential for 5G and beyond communications in order to support hyper-dense ultra-fast access networks [3]. Furthermore, with the development of the Internet of Things (IoT) concept, the rate at which physical devices are connected to the internet is increasing exponentially [7]–[11]. Currently, radio frequency (RF) is widely used across various wireless applications. However, in relation to meeting the growing demand for 5G wireless capacity and serving the IoT paradigm, the currently used RF spectrum is insufficient. The electromagnetic spectrum, with favorable communication properties below 10 GHz, is widely used by existing wireless technologies and has almost been exhausted; therefore, it is predicted that the massive connectivity demand of future mobile data traffic will not be met by existing wireless technologies [12]. Moreover, this band (below 10 GHz) has limitations such as a small spectrum band, regulations related to spectrum use, and severe interference among nearby RF access points. Therefore, researchers are looking for new complementary spectra, such as millimeter and nanometer waves, for wireless communication connectivity [13]. At the World Radio Conference 2015 (WRC-15), the International Telecommunication Union (ITU) proposed 11 new candidate bands for International Mobile Telecommunication-2020 (IMT-2020), i.e., 5G communication [11].

The RF band lies between 3 kHz and 300 GHz of the electromagnetic spectrum. The use of this band is strictly regulated by local and international authorities. In most cases, RF sub-bands are entirely licensed to certain operators, e.g., cellular phone operators, television broadcasters, and point-to-point microwave links [14]. The optical spectrum is considered as a promising solution for the development of future high-density and high-capacity networks. Wireless connectivity based on the optical spectrum is termed as optical wireless communication (OWC). In comparison with RF-based networks, OWC-based network technologies offer unique advantages. OWC systems can provide high-data-rate services for communication distances ranging from a few nanometers to more than 10,000 km. It can perform well both for indoor and outdoor services. However, OWC systems suffer owing to their sensitivity to blocking by obstacles and to their limited transmitted power. Therefore, the coexistence of OWC and RF systems may provide an effective solution for the huge demands of upcoming 5G and beyond communication systems.

### A. What are the OWC Technologies?

The term OWC [14]–[20] refers to optical transmission in which guided visible light (VL), infrared (IR), or ultraviolet (UV) spectrum are used as propagation media. OWC systems operating in the VL band are commonly categorized as visible light communication (VLC). VLC can simultaneously offer communication, illumination, and localization [21]. Terrestrial point-to-point free space optical (FSO) communication systems [14], [22] are operated at IR, VL, and UV frequencies. UV communication can provide high-data-rate non-line-of-sight (NLOS) and line-of-sight (LOS) optical communication links [23]–[25].

In this paper, we review the development of different OWC technologies such as VLC, light fidelity (LiFi), optical camera communication (OCC), FSO communication (FSOC), and light detection and ranging (LiDAR). Several OWC technologies are being developed to meet the demand of 5G and beyond communication systems. A few of these technologies have already been commercialized and others are in the developmental stage. Light-emitting diode (LED) luminaires or laser diodes (LDs) are used for high-speed data transfer in VLC [26]–[28]. LiFi [29]–[31], a technology complementing wireless fidelity (WiFi), is an OWC system that provides high-speed wireless communication along with illumination. The communication purposes served by these technologies are similar. LiFi technology normally uses LEDs as transmitters and photodetectors (PDs) as receivers. However, combined LDs with an optical diffuser [31]–[33] can be also used as transmitters and light sources. VLC and LiFi can provide high-data-rate communication for indoor applications but they are not very effective for outdoor applications. Furthermore, they cannot provide long-distance communication. OCC [34]–[36], a subsystem of OWC, uses LEDs as the transmitter and a camera or image sensor (IS) as the receiver. OCC can provide noninterference communication and a high signal-to-noise ratio (SNR) even in outdoor environments [35]. Moreover, stable performance is achievable even when the communication distance increases [36]. FSO communication [37]–[41] mainly uses the near IR (NIR) band as the communication medium. However, the UV and VL bands also fall into the FSO category. In FSO systems, narrow beams of intense light are used to establish a high-data-rate communication link between two ends within the range of inter-chip to inter-satellite connections [37]–[41]. LiDAR [42], [43] is an optical remote sensing technology for very-high-resolution 3D mapping. It analyses the properties of scattered light to find the range and/or other information related to a distant target.

### B. OWC Application Platforms

OWC has a wide range of applications [12]–[26], [29]–[42]. Fig. 1 shows several areas where OWC can be applied, such as industry, healthcare, railway stations, transportation, homes, offices, shopping malls, underwater, and space. For these application platforms, all types of communication, such as device-to-device (D2D); machine-to-machine (M2M); chip-to-chip; device/machine-to-user; user-to-device/machine; vehicle-to-infrastructure, vehicle-to-vehicle, and infrastructure-to-vehicle (V2X); point-to-point; multipoint-to-point; and point-to-multipoint, can be performed using OWC technologies. Based on the application type, required data speed, and platform, various OWC technologies can be applied.

### C. OWC for 5G and IoT

OWC has become a promising technology for supporting high-data-rate 5G communication and the massive connectivity of IoT. 5G and beyond communication systems must possess the necessary features for integrating ultra-dense heterogeneous networks. A straightforward but extremely effective way to increase network capacity is to make the cells smaller [6]. Networks are now rapidly being developed to include nested small cells such as picocells and femtocells for 5G and beyond communications. VLC, LiFi, and OCC can provide ultra-dense small cell hotspot services to meet the demands of 5G. Furthermore, FSO, LiFi, and VLC can effectively provide high-capacity backhaul support for 5G and beyond communication systems. OWC technologies have very low power consumption, which is a key requirement of 5G [2]. A reliable connection, which is the main priority for 5G communication, can also be provided using OWC technology. It can provide secure communications, as required by 5G, and can also connect a large and diverse set of 5G devices for indoor and outdoor communications. It will offer extreme densification and offloading to improve the area spectral efficiency. To support the IoT paradigm, OWC technology is able to provide massive connectivity through low power LED technologies.

### D. Related Review Literatures

The use of optical bands complementary to RF will relieve the problems caused by spectrum shortage in RF-based wireless communication. Globally, researchers are attempting to adapt many applications currently based on RF communication to OWC. A comparative study among the OWC technologies is very essential to understand those. There exist few literature reviews [14]–[16], [18], [20], [44]–[48] in different aspects of OWC. However, a comparative study on optical wireless technologies is missing yet. Reference [14] provides an overview of OWC, highlighting only the advantages and range of application areas for this technology. An overview of OWC systems, mainly focusing on the principle of operation for VLC, FSO, transcutaneous OWC, underwater OWC, and optical scattering communications is found in [15]. The authors of [16] discuss the various standards for OWC. In [18], the authors present an overview of OWC technologies, emphasizing their deployment in communication systems. Reference [20] presents an overview of indoor OWC systems, mainly focusing on the challenges, research issues, and prospects of such systems. The authors of [44] present the trends towards wireless-optical convergence. It mainly discusses integration of wireless and optical networks and dynamic bandwidth allocation. Reference [45] presents the potential for integration of FSO and VLC technologies as a single field of study. It provides the key challenges involved in designing technologies jointly performing the lighting and networking functions. In [46], the authors provide a survey on VLC with an emphasis on challenges faced in indoor applications. It discusses the modulation schemes, dimming techniques, filtering, equalization, compensation, and beamforming for indoor VLC. The authors of [47] review the key enabling access technologies and progress advancements of optical broadband network and wireless access technologies. The book chapters of [48], mainly discuss modulation concepts, MIMO technique, channel modeling, and capacity analysis for OWC.

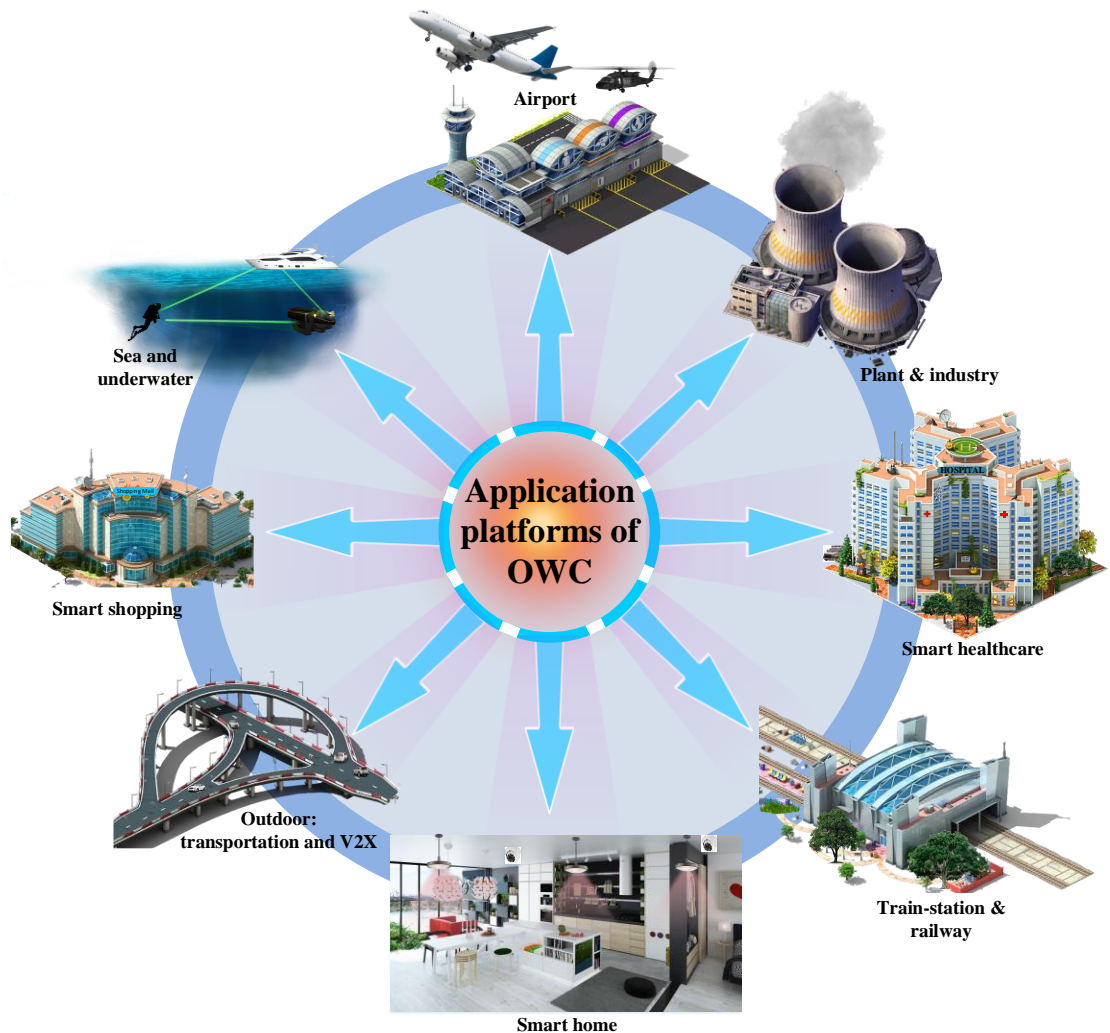

Fig. 1. Example of OWC application platforms.

### E. Contribution of This Paper

In this survey, we present a general overview and comparison of emerging OWC technologies. The main focus of this paper is to make clear the differences among different optical wireless technologies and between promising optical wireless technologies and their respective similar existing RF technologies. Here we discuss OWC systems from various viewpoints, e.g., the types of optical sub-bands used

(IR/VL/UV); the types of technology applied (LiFi/VLC/OCC/FSO/LiDAR); the application scenarios (range of communication distance, data rate/accuracy/reliability, LOS/NLOS, indoor/outdoor, and underwater/vehicular/space); the types of transmitter/receiver (LED/LD and PD/camera); the network architecture; the principle of operation; and the types of application (illumination/localization/communication/high-speed link/ imaging/mapping), to highlight the differences among the technologies.

The rest of the paper is organized as follows. Section II provides an overview of OWC systems; the classification of OWC based on technology, communication distance, and communication media are presented in this section. Different OWC technologies, their applications, and principles of operation are presented in Section III. Section IV summarizes the review and discusses the future direction of research issues. The survey is concluded in Section V. For ease of reference, the various abbreviations used in this paper are summarized in Table I.

TABLE I
LIST OF ACRONYMS

| 5G | Fifth-Generation |
|---|---|
| CDMA | Code-Division Multiple-Access |
| CN | Core Network |
| CSK | Color Shift Modulation |
| D2D | Device-to-Device |
| EHF | Extremely High Frequency |
| EUV | Extreme Ultra Violet |
| FIR | Far Infrared |
| FOV | Field of View |
| FSOC | Free Space Optical Communication |
| FSO | Free Space Optical |
| HF | High Frequency |
| ID | Identification |
| IM/DD | Intensity Modulation and Direct Detection |
| IMT | International Mobile Telecommunication |
| IoT | Internet of Things |
| IR | Infrared |
| IS | Image Sensor |
| ITU | International Telecommunication Union |
| LD | Laser Diode |
| LiDAR | Light Detection and Ranging |
| LF | Low Frequency |
| LiFi | Light Fidelity |
| LED | Light-Emitting Diode |
| LOS | Line-of-Sight |
| LWIR | Long-Wavelength Infrared |
| M2M | Machine-to-Machine |
| MF | Medium Frequency |
| MWIR | Mid-Wavelength Infrared |
| NIR | Near Infrared |
| NLOS | Non-Line-of-Sight |
| NUV | Near Ultraviolet |
| OCC | Optical Camera Communication |
| OCI | Optical Communication Image Sensor |
| OFDM | Orthogonal Frequency Division Modulation |
| OOK | On-Off Keying |
| OWC | Optical Wireless Communication |
| PD | Photodetector |
| PM | Pulse Modulation |
| RADAR | Radio Detection and Ranging |
| RF | Radio Frequency |
| SHF | Super-High Frequency |
| SNR | Signal-to-Noise Ratio |
| SWIR | Short-Wavelength Infrared |
| TG | Task Group |
| TIR | Thermal Infrared |
| UE | User Equipment |
| UHF | Ultra-High Frequency |
| UV | Ultraviolet |
| UVA | Ultraviolet A |
| UVB | Ultraviolet B |
| UVC | Ultraviolet C |
| VCSEL | Vertical-Cavity Surface-Emitting Laser |
| VHF | Very High Frequency |
| VL | Visible Light |
| VLC | Visible Light Communication |
| VLF | Very Low Frequency |
| WBAN | Wireless Body Area Network |
| WiFi | Wireless Fidelity |
| WRC | World Radio Conference |

## II. OVERVIEW

The electromagnetic spectrum consists wide range of various frequency bands. Currently only few bands are used for communication purposes. Wide range of electromagnetic spectrum opened the opportunity to extend the use of frequency band for communication purposes. Table II shows the detail classification of electromagnetic spectrum. A huge portion of the electromagnetic spectrum i.e., optical spectrum is available for OWC systems. Therefore, OWC systems are likely a promising solution for the future heterogeneous network as a complementary option to RF. Several researchers have already demonstrated high data rates by using OWC [12], [49]–[51]. Each of the optical bands IR, VL, and UV has unique advantages and limitations. Based on these optical bands, various wireless technologies are being developed. The wavelengths of NIR and VL are nearly close and they exhibit qualitatively similar behavior from the viewpoint of communication [52]. However, IR is invisible to humans and can be used for applications where illumination is not needed. UV can provide high-data-rate LOS communication over short and long distances, and NLOS communications over medium and long distances [23]. The most important advantage of OWC is the wide availability of incoherent front-end components that allow for relatively straightforward realization of intensity modulation and direct detection (IM/DD) [12]. Because of this IM/DD approach, OWC is based on baseband modulation techniques for which a large variety of off-the-shelf high-bandwidth components are available already. A few important advantages of wireless communication by using optical spectrum over the other spectrum regions include [12], [21] (i) huge unregulated bandwidth; (ii) high level of security (because light signals cannot pass through walls and can therefore be confined within a well-defined coverage zone); (iii) low power consumption; (iv) low cost; (v) no interference with electronic equipment and other communication networks; (vi) very high achievable SNR; and (vii) easy integration into the existing lighting arrangement. On the other hand, OWC systems have few limitations such as (i) obstacles block transmission channel; (ii) sensitive to sudden blocking a connection; and (iii) limited transmitted power.

### A. OWC Classification

Based on the communication distance, OWC systems can be classified into five categories, namely, ultra-short range, short range, medium range, long range, and ultra-long range [37], [52]–[63].

- Ultra-short range OWC: In this category of OWC, nm/mm-level communications are performed. An example of this type of communication is nm distance chip-to-chip communication [53]–[56].
- Short range OWC: Wireless body area network (WBAN), wireless personal area network (WPAN), and underwater communications are few examples of this category [57], [58], [63].
- Medium range OWC: This communication range comprises VLC-based WLANs and outdoor V2X communications [52], [59], [60].
- Long range OWC: This communication range provides km-range of communication, for example, inter-building connections [37].
- Ultra-long range OWC: This communication range comprises inter-satellite, satellite-earth, satellite-to-airplane, airplane-to-satellite, airplane-to-airplane, and airplane-to-ground links [61], [62].

TABLE II
ELECTROMAGNETIC SPECTRUM [14], [15], [22], [23], [26], [29], [37], [64], [65]

| Spectral category/sub-category | | | Frequency | Wavelength |
|---|---|---|---|---|
| Radio wave | Very low frequency (VLF) | | 3 - 30 kHz | 100 - 10 km |
| | Low frequency (LF) | | 30 - 300 kHz | 10 - 1 km |
| | Medium frequency (MF) | | 0.3 - 3 MHz | 1000 -100 m |
| | High frequency (HF) | | 3 - 30 MHz | 100 – 10 m |
| | Very high frequency (VHF) | | 30 - 300 MHz | 10 -1 m |
| | Ultra-high frequency (UHF) | | 0.3 - 3 GHz | 1 – 0.1 m |
| | Super-high frequency (SHF) | | 3 - 30 GHz | 100 - 10 mm |
| | Extremely high frequency (EHF)/Millimeter wave | | 30 - 300 GHz | 10 - 1 mm |
| | Microwave | P-Band | 0.225 - 0.39 GHz | 1330 - 769 mm |
| | | L-Band | 0.39 - 1.55 GHz | 769 - 193 mm |
| | | S-Band | 1.55 - 5.2 GHz | 193 - 57.7 mm |
| | | C-Band | 3.9 - 6.2 GHz | 76.9 - 48.4 mm |
| | | X-Band | 5.2 - 10.9 GHz | 57.7 - 27.5 mm |
| | | Ku-Band | 12 - 18 GHz | 25 - 16.67 mm |
| | | K-Band | 10.9 - 36 GHz | 27.5 - 8.33 mm |
| | | Q-Band | 36 - 46 GHz | 8.33 - 6.52 mm |
| | | V-Band | 46 - 56 GHz | 6.52 - 5.35 mm |
| | | W-Band | 56 - 100 GHz | 5.35 - 3 mm |
| Optical | IR | Far infrared (FIR) | 0.3 - 20 THz | 1-0.015 mm |
| | | Thermal infrared (TIR) — Long-wavelength infrared (LWIR) | 20 - 37.5 THz | 0.015-0.008 mm |
| | | Thermal infrared (TIR) — Mid-wavelength infrared (MWIR) | 37 - 100 THz | 0.008-0.003 mm |
| | | Short-wavelength infrared (SWIR) | 100 – 214.3 THz | 3000000– 1400 nm |
| | | Near infrared (NIR) | 214.3 - 394.7 THz | 1400-760 nm |
| | Visible light | Red | 394.7 - 491.8 THz | 760 - 610 nm |
| | | Orange | 491.8 - 507.6 THz | 610 - 591 nm |
| | | Yellow | 507.6 - 526.3 THz | 591 - 570 nm |
| | | Green | 526.3 - 600 THz | 570 - 500 nm |
| | | Blue | 600 - 666.7 THz | 500 - 450 nm |
| | | Violet | 666.7 - 833.3 THz | 450 - 360 nm |
| | UV | Ultraviolet A (UVA) | 750 - 952.4 THz | 400 - 315 nm |
| | | Ultraviolet B (UVB) | 952.4 - 1071 THz | 315 - 280 nm |
| | | Ultraviolet C (UVC) | 1.071 - 3 PHz | 280 - 100 nm |
| | | Near ultraviolet (NUV) | 0.750 - 1 PHz | 400 - 300 nm |
| | | Middle ultraviolet | 1 - 1.5 PHz | 300 - 200 nm |
| | | Far ultraviolet | 1.5 - 2.459 PHz | 200 - 122 nm |
| | | Hydrogen Lyman-alpha | 2.459 - 2.479 PHz | 122 - 121 nm |
| | | Extreme ultra violet (EUV) | 2.479 - 30 PHz | 121 - 10 nm |
| | | Vacuum ultraviolet | 1.5 - 30 PHz | 200 - 10 nm |
| X-ray | | Soft X-ray | 30 - 3000 PHz | 10 - 0.1 nm |
| | | Hard X-ray | 3 - 300 EHz | 100 - 1 pm |
| Gamma ray/ Cosmic ray | | | 300 - 30000 EHz | 1000 -10 fm |

Fig. 2 shows the classification of OWC applications based on transmission range. OWC can be used for WBAN, WPAN, wireless metropolitan area network (WMAN), wide area network (WAN), etc. Fig. 3 shows the applications for OWC systems for different ranges of communication. It shows that a wide range of communications is possible using OWC, i.e., from ultra-short range inter-chip up to ultra-long range inter-satellite communications.

### B. Infrared vs VL vs Ultraviolet for OWC

Based on the transmission spectrum, OWC can be studied in three categories, namely, infrared, VL, and ultraviolet. The VL spectrum is widely used in VLC, LiFi, and OCC technologies. In limited cases, the VL spectrum is also used in FSO. Terrestrial point-to-point FSO systems are widely operated in the NIR spectrum. NIR is also widely used for OCC, LiFi, and LiDAR technologies. There has been a notable interest in UV communication using solar-blind UV spectrum (200–280 nm) for its capability of establishing NLOS optical communication links. High-speed NLOS and LOS communications can be achieved using UV communication [23], [24]. LiFi and FSO can also use UV spectrum as a communication medium. Table III shows a comparison of IR, VL, and UV bands from the viewpoint of their use in OWC technologies.

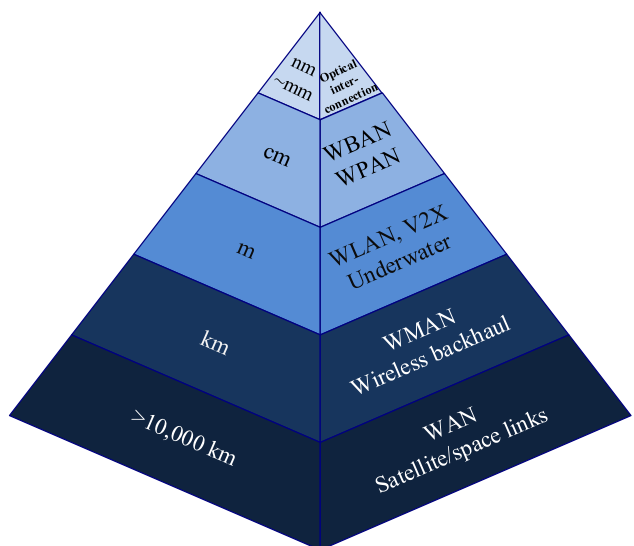

Fig. 2. Classification of OWC applications based on transmission range.

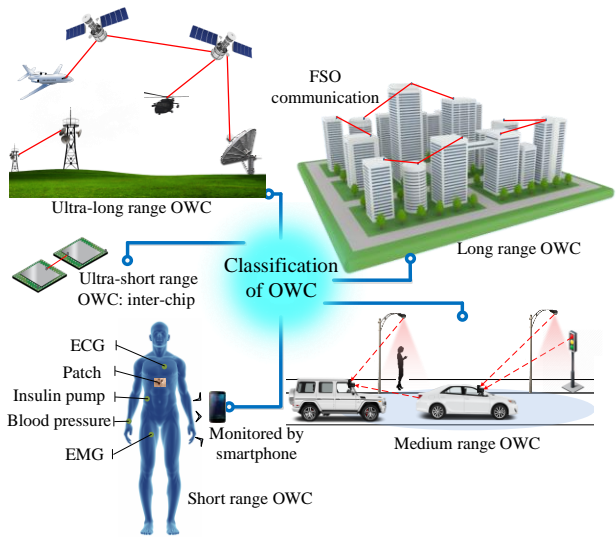

Fig. 3. Examples of OWC for different ranges of communications.

TABLE III
OWC TECHNOLOGIES USING IR/VL/UV

| Issue | IR | VL | UV |
|---|---|---|---|
| Wavelength | 760 nm-1mm | 360-760 nm | 10-400 nm |
| Communication distance | • Short and medium ranges in LiFi and OCC<br>• Ultra-short, short, medium, long, and ultra-long ranges in FSOC<br>• Short, medium, long, and ultra-long ranges in LiDAR | • Short and medium ranges in VLC, LiFi, and OCC<br>• Ultra-short, short, medium, long, and ultra-long ranges in FSOC<br>• Short , medium, long, and ultra-long ranges in LiDAR | • Short and medium ranges in LiFi<br>• Ultra-short, short, medium, long, and ultra-long ranges in FSOC |
| Advantage | Not visible for human eye considering the cases where illumination is not important | • Safe for human<br>• It can be used for illumination and communication purposes simultaneously | • Not visible for human eye considering the cases where illumination is not important<br>• High data rate NLOS communication is possible |
| Limitation | • Not always safe for human<br>• LOS communication and very limited low data rate NLOS communication using reflection of IR | • LOS communication and very limited low data rate NLOS communication using reflection of light<br>• Visibility of light when illumination is not required | Not safe for human |
| Communication Technologies | LiFi, OCC, FSO, and LiDAR | VLC, LiFi, OCC, FSO, and LiDAR | LiFi and FSO |
| Illumination | No | Communication with and without illumination | No |

### C. NLOS Communications

OWC is widely used for LOS communications. However, NLOS communication is also possible using OWC technologies. High-data-rate NLOS communication can be achieved using UV spectrum. Fig. 4 shows NLOS communications using IR, VL, and UV. Fig. 4(a) shows a low-data-rate NLOS communication system achieved using the reflection of IR and VL. A high-data-rate NLOS communication system is shown in Fig. 4 (b). To perform NLOS UV communication, an omnidirectional transmitter is directed at an elevation, while the receiver is required to intersect the transmitted beam. The transmitter comprises a xenon flashtube inside a searchlight dish. By design, the front-end receiver of such a system must have high gain and low noise, and it is built using a photo-multiplier tube and a counter [66]. The atmospheric gases ozone and oxygen are strong absorbers of light in the UV spectral region [67]. A filter in front of the phototube generates a pulse when it detects an incident photon of operation. The knowledge of the transmitted and received beam positions is important to achieve higher data rate, reduced latency, and reduced transmitted power.

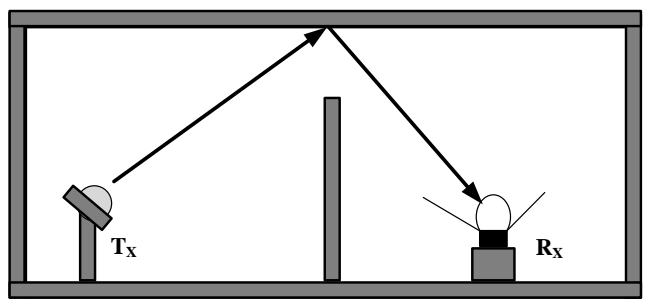

(a)

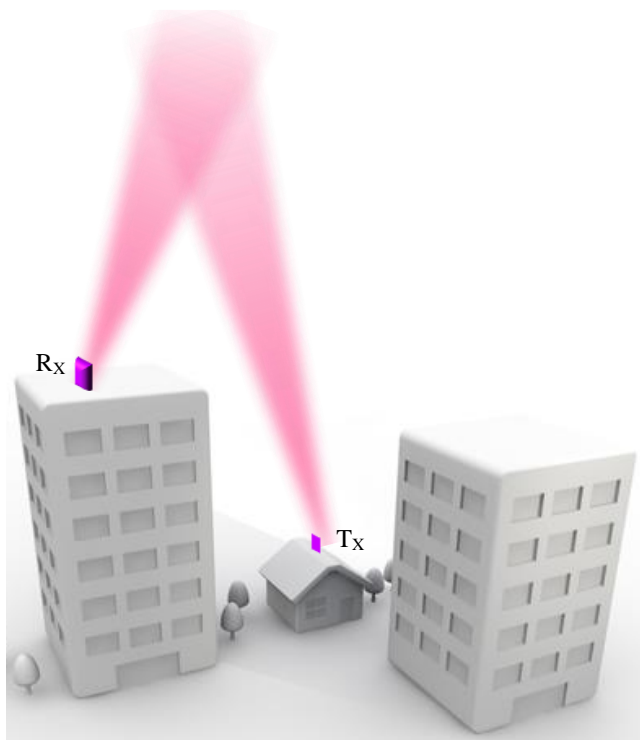

(b)

Fig. 4. NLOS communications, (a) NLOS communication using IR and VL, (b) NLOS communication using UV.

### D. LED vs LD for OWC

LEDs and LDs offer unique advantages and limitations from the viewpoint of their use in OWC systems. The advantages and limitations of OWC systems based on LEDs and LDs are briefly listed below:

**Advantages of LEDs in OWC:** Owing to recent advancements in solid-state lighting, there has been a trend in the last decade to replace incandescent and fluorescent lamps with high-intensity solid-state white LEDs. LEDs have benefits such as extremely high energy efficiency, longer lifespan, compact form factor, lower heat generation, reduced usage of harmful materials in design, and improved color interpreting without using harmful chemicals [26]. Because of these benefits, LED adoption has been rising consistently, and it is expected that nearly 75% of all illumination will be provided by LEDs by the year 2030 [68]. One additional important benefit of LEDs is that they are capable of switching to different light intensities very rapidly. This special functionality of high switching rate creates the opportunity to use LEDs as OWC transmitters for high-speed communication and highly efficient lighting source simultaneously. Therefore, LEDs can serve the dual purposes providing of highly efficient illumination and very-high-speed communication.

**Limitation of LEDs in OWC:** All light produced by an LED is incoherent in nature. Therefore, all the waves are not in phase and the transmitted optical power produced by an LED is comparatively very low. Also natural and artificial light sources cause interference for an LED source light.

**Advantages of LDs:** Lasers are monochromatic. Therefore, all light produced using a laser is of a single, specific wavelength. Coherent lights mean all individual light waves are precisely lined up with each other, all waves travel in the same direction, in the same manner, and at the same time [69]. Laser lights are concentrated and directed forward because of coherent characteristics. Therefore, laser light can travel long distances. Moreover, given this nature, LDs cause less interference and provide a high-data-rate compared to LED lights.

**Disadvantages of LDs:** LD has low aperture. Only point-to-point communication is possible using LD.

The idea of using LD-based FSO communication has been well known for many years. High-data-rate communication has already been demonstrated for mobile access using IR LDs [70]. Nevertheless, LDs are not popular due to possible health threats (e.g., eye injury, hyperthermia, coagulation, and ablation due to thermal effects of laser radiation [71], [72]), cost, color mixing complication, and the questionable quality of laser light for illumination purposes [33]. A recent study shows that diffused laser light does not compromise the user experience compared with conventional light luminaries [73].

### III. ENABLING TECHNOLOGIES

Each of the optical technologies (VLC, OCC, LiFi, FSO, and LiDAR) has a unique architecture and principle of operation. They may also differ in terms of modulation technique, transmitting system, receiving system, and communication media. Fig. 5 shows the basic architectures of the various OWC technologies. Different OWC technologies use either LEDs or LDs as physical transmitters. PDs or cameras (or ISs) are used as physical receivers, and IR, VL, or UV spectra are used as communication media. This figure explains how the various OWC technologies differ from each other in terms of transmitter, receiver, and communication medium.

### A. Visible Light Communication (VLC)

VLC, a subset of OWC, has emerged as a promising technology in the past decade. VLC based on LEDs or LDs can be a promising solution for upcoming high-density and high-capacity 5G wireless networks. IoT is becoming increasingly important  because it allows a large number of

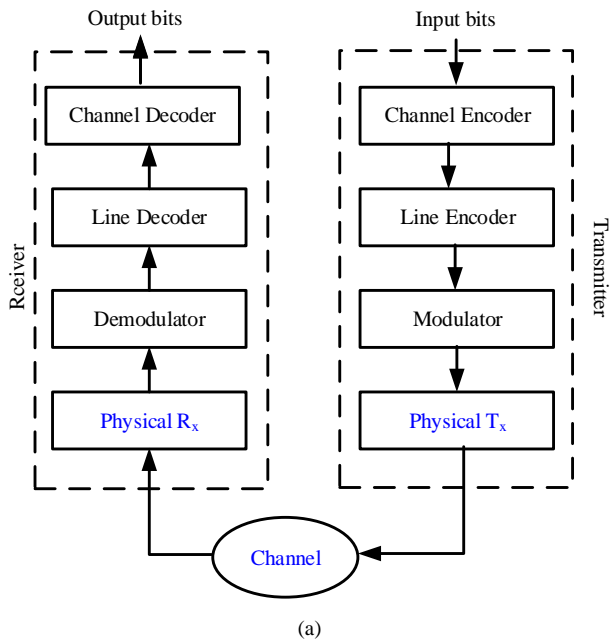

(a)

| | Channel | Physical T$_x$ | Physical R$_x$ |
|---|---|---|---|
| OWC | IR/VL/UV | LED/LD | PD/Camera |
| VLC | VL | LED/LD | PD/Camera* |
| LiFi | VL | LED/LD | PD |
| OCC | IR/VL | LED | Camera |
| FSO | IR/VL/UV | LD | PD |
| LiDAR | IR/VL | LD | None (only reflected at object) |

*Camera (or IS) receiver used for VLC is also categorized as OCC

(b)

| | Channel | Physical T$_x$ | Physical R$_x$ |
|---|---|---|---|
| OWC | IR/VL/UV | LED/LD | PD/Camera (or IS) |
| VLC | VL | LED/LD | PD |
| LiFi | IR/VL/UV | LED* | PD |
| OCC | IR/VL | LED | PD/Camera |
| FSO | IR/VL/UV | LD | PD |
| LiDAR | IR/VL | None (only reflected from object) | PD/Camera |

*Combined LDs with optical diffuser can also be used

(c)

Fig. 5. Basic architectures of different OWC technologies, (a) basic block diagram, (b) downlink communication, (c) uplink communication.

devices to be connected for sensing, monitoring, and resource sharing, and VLC could play an important role to this end [74]. VLC technology offers 10,000 times more bandwidth capacity than RF-based technologies [75]. VLC based on the VL spectrum provides significant opportunities for application in homes, offices, cars, airplanes, trains, and roadside [21]. Moreover, it is not harmful to humans. However, this technology suffers from a few limitations: (i) it is not very effective for outdoor applications; and (ii) it cannot perform long-distance communication.

LEDs can switch to different light intensity levels at a very fast rate, which allows data to be modulated through LED at a speed that the human eye cannot detect [65], [76], [77]. Therefore, these energy-efficient LED devices can be used for illumination, communication, and localization [78], a unique feature not available in other devices. Potential VLC data rates of over 10 Gbps have already been demonstrated using LEDs [12]. A data rate of 3 Gbps has been achieved using a single color LED [51]. The main disadvantage of LEDs is their inherent tradeoff between bandwidth and optical efficiency. Therefore, LDs can be considered as a promising alternative for better utilization of the VL spectrum for communication purposes. A data rate of 100 Gbps is achievable at standard indoor illumination levels using LD-based VLC [12]. VLC has many indoor and outdoor applications, a few of which are as follows [26]–[28], [79]:

- Smart lighting: Lighting using the VLC arrangement facilitates illumination, communication, and control simultaneously. Therefore, it can provide considerable savings in terms of cost and energy consumption.
- Wireless connectivity: We can create very-high-speed wireless connectivity with inherent security by using VLC.
- V2X communication: Many cars already have LED lamps. Traffic lights, traffic signage, and street lamps are adopting LED technology. Therefore, VLC creates a great opportunity to create secure and high-data-rate communication links for vehicle-to-vehicle and vehicle-to-infrastructure connectivity.
- Hospitals: Conventional RF-based communications are not desirable in many parts of hospitals, especially around MRI scanners and in operating theaters. Therefore, the use of VLC is a good alternative to RF-based communication in hospitals.
- Spectrum relief: RF is very much congested, and therefore, data can be offloaded using VLC.
- Aviation: RF is undesirable for use by passengers of aircraft. LEDs are already used in aircraft for illumination and can also be used instead of wires to provide media services to passengers. This reduces the cost of aircraft construction and its weight.
- Underwater communications: RF-based communication does not work underwater, but VLC can support high-speed data transfer in this environment.
- Smart displaying signboards: Nowadays signboards are used everywhere, including shopping malls, roads, airports, and bus stops. These displays are made using arrays of LEDs. These LEDs can convey information through VLC technology.
- Location-based services: Each LED source can be identified uniquely, so location-based services can be identified quickly and accurately by using VLC.
- Local area networks: LED-based VLC can be used to create high-speed local area networks.
- Sound communication system: Different colored LEDs can be used for the transmission of sound [80].

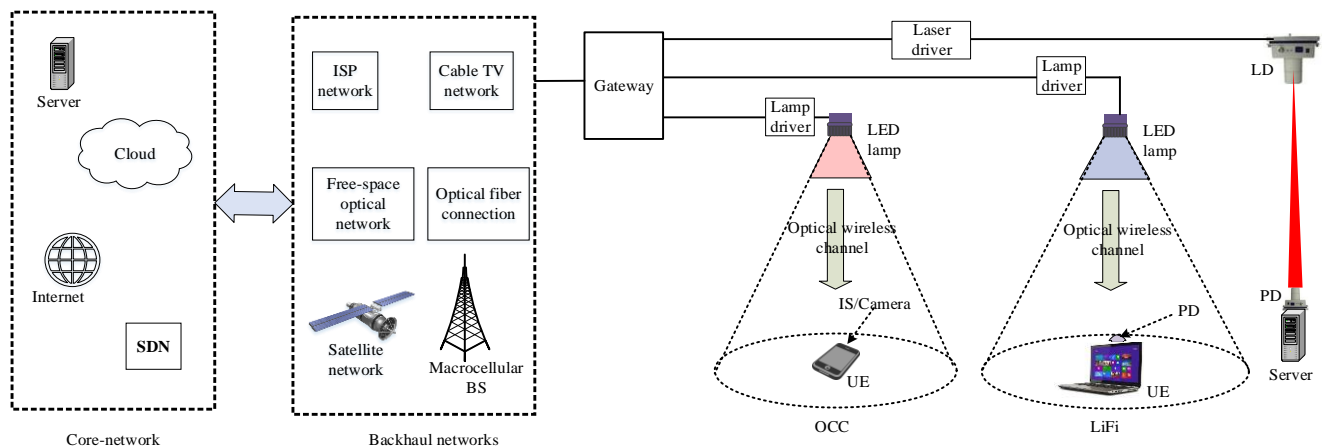

Fig. 6. Basic architecture of UE-to-CN connectivity in VLC system.

A basic architecture for user equipment (UE)-to-core network (CN) connectivity in VLC system is shown in Fig. 6. LED lamps or LDs are used as transmitters. Cameras or ISs or PDs are used as receivers. Then, VLC access networks are connected to the Internet, any server, or CN through wired or wireless backhaul connectivity. The backhaul network connectivity can be either internet service provider (ISP) or cable TV or FSO or optical fiber or cellular or even satellite. In indoor environments, it is normally a wired connection. However, in outdoor cases, especially in vehicular environments, wireless backhaul connectivity is used. LiFi and OCC also use similar connectivity. The only differences are the physical transmitter, receiver, and transmission media.

## B. Light Fidelity (LiFi)

LiFi can be classified as a nm-wave communication [29]. The main differences between LiFi and VLC are (i) VLC uses only the VL portion of the light spectrum, whereas LiFi uses VL in the forward path but it is possible to use either IR, VL, or UV in the return path [29]–[31], [80]–[82]; (ii) VLC systems can be either unidirectional [83] or bidirectional [84], whereas LiFi is a bidirectional communication system [31]; (iii) LiFi must provide seamless user mobility, whereas mobility support is not mandatory for VLC [31]; and (iv) VLC systems include any of point-to-point, point-to-multipoint, and multipoint-to-point communications, whereas LiFi systems must include multiuser communications, i.e., point-to-multipoint and multipoint-to-point communications. Hence, LiFi is a complete wireless networking system and supports seamless user mobility; therefore, it forms a new small attocell layer within existing heterogeneous wireless networks. As such, a VLC system will be treated as LiFi only if it has LiFi features (e.g., multiuser communication, point-to-multipoint and multipoint-to-point communications, and seamless user mobility). On the other hand, a LiFi system can be treated as VLC only when VL is used as the transmission media.

The term LiFi is similar to WiFi, with the exception that optical spectrum rather than RF spectrum is used for transmission. Existing LEDs can be used as the transmitter for LiFi cellular deployment [29]. Moreover, LDs combined with optical diffuser can also be used as LiFi transmitters. Speeds of over 3 Gbps from just a single microLED have been demonstrated [33], and a speed of 56 Gbps was achieved using a vertical-cavity surface-emitting laser (VCSEL) [30]. The LiFi attocells concept can reduce cell size compared to mm-wave communication [85]. Large amounts of data can be offloaded to LiFi attocells from congested RF-based networks. The lower-tier LiFi attocells within existing multi-tier heterogeneous wireless networks have zero interference from and add zero interference to RF-based networks such as femtocell networks [29]. Fig. 7 shows selected application scenarios (both indoor and outdoor) of VLC/LiFi technologies. The application scenarios of VLC and LiFi are similar. In the indoor environment, communication and illumination are performed simultaneously. For certain cases in the outdoor environment, both communication and illumination are performed (e.g., in street lamps and vehicle headlamps), whereas in other cases, only communication is performed (e.g., in traffic signals and the rear lamps of vehicle). The examples shown in Fig. 7(a) apply to VLC as well as LiFi communication, depending on the supported communication media for uplink and the provided features. In Fig. 7(b), street light–smartphone communication is an example including both VLC and LiFi systems. On the other hand, infrastructure-to-vehicle, vehicle-to-infrastructure, and vehicle-to-vehicle are examples of VLC.

The potential advantages of LiFi over WiFi include very fast data rate, lower cost, readily available spectrum capacity, and better security. The security features of LiFi make it promising for use in electromagnetic-sensitive areas such as hospitals, nuclear plants, and aircraft cabins. However, LiFi technology is not very effective for outdoor applications and also it cannot offer long-range communications. A few important differences between LiFi and WiFi are listed in Table IV.

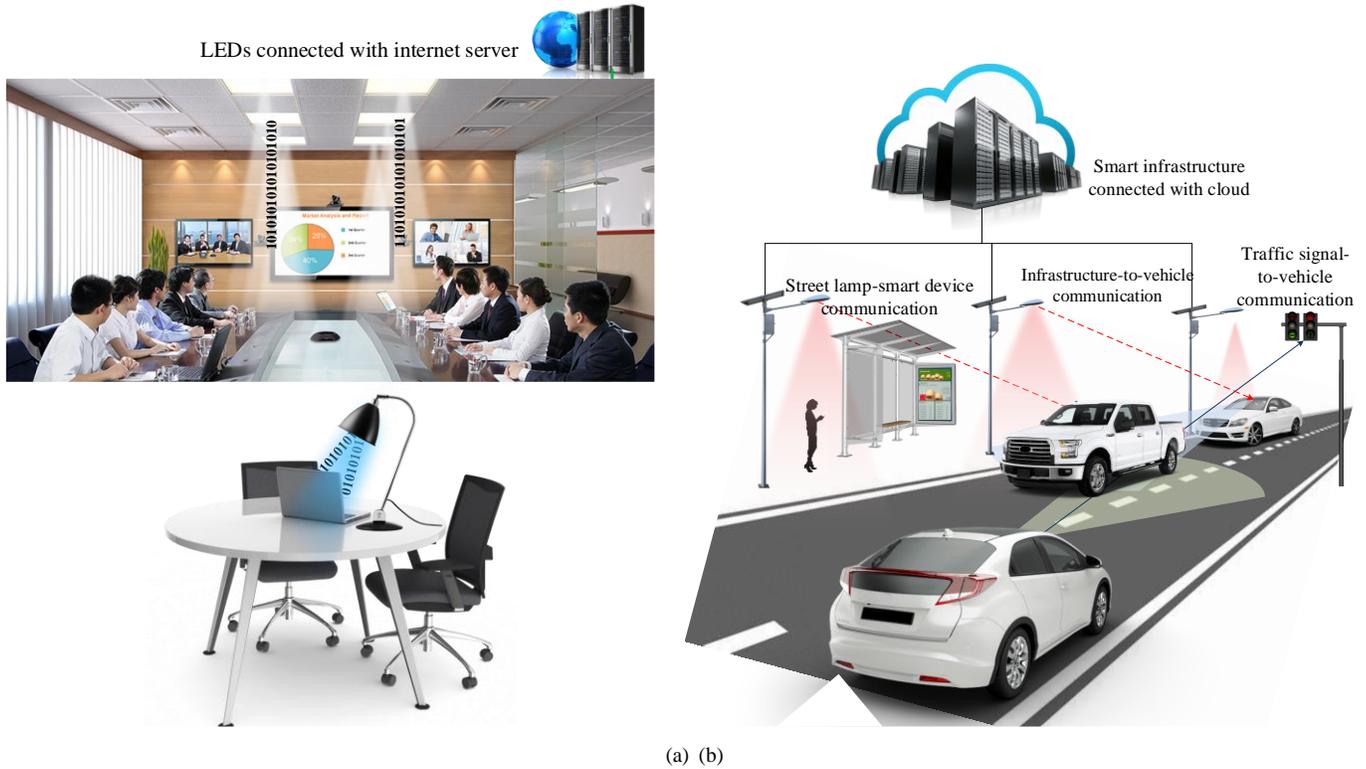

(a)  (b)

Fig. 7. Selected application scenarios of VLC/LiFi systems, (a) indoor environment, (b) outdoor environment.

TABLE IV
DIFFERENCES BETWEEN LiFi and WiFi [12], [15], [26], [33]

| Issue | LiFi | WiFi |
|---|---|---|
| Transmission and reception medium | VL for downlink and VL or IR or UV for uplink | RF waves |
| Maximum data rate | 10 Gbps using LED and 100 Gbps using LD (combined with optical diffuser) | 6 Gbps using IEEE 802.11ad |
| Communication distance | 10 m | 100 m |
| Interference level | Low | High |
| Transition | Directional | Omnidirectional |
| IEEE standard | 802.15.7m | 802.11 |
| Infrastructure cost | Less | More |
| Modulation technique | Direct Current biased Optical Orthogonal Frequency Division Multiplexing (DCO-OFDM) | Direct Sequence Spread spectrum (DSSS) |
| Power consumption for communication purpose | Very small | Comparatively high |
| Human harmness | No | Yes [86] |

## C. Optical Camera Communication (OCC)

Currently, we are highly accustomed to the presence of smart devices in our daily life. The majority of these devices are equipped with front and/or rear cameras with LED flashes. This facilitates OWC implementation using devices that use a flash and a camera as the transceiver pair. This implementation does not require any further hardware modification. Therefore, there is growing interest in OWC implementation using LEDs, displays, or other light sources as the transmitter and a camera (or IS) as the receiving module. The main differences between OCC and VLC are (i) the VLC system must use VL as a communication medium, whereas the OCC system can use both IR and VL; (ii) the VLC system can use both LEDs and LDs as physical transmitters, whereas the OCC system can use only LEDs; and (iii) a PD or camera (or IS) can be used as the receiver in VLC, whereas a camera (or IS) is used as the receiver in OCC. Therefore, a VLC system can be treated as an OCC system when it uses LEDs and a camera (or IS) as its transmitter and receiver, respectively. In addition, an OCC system can be considered a VLC when it uses VL as the communication medium.

Although considerable research effort has been focused on the development of indoor VLC systems, there remains a need for research on outdoor VLC systems. There are many challenges [21], [87]–[91] associated with LED/PD-based VLC/LiFi systems, including (i) short communication distance; (ii) drastic attenuation of received signal strength; (iii) interference due to artificial and natural light sources; (iv) noise due to ambient light radiation from the sun, skylights, streetlights, and other sources in outdoor environments; (iv) the large amount of noise from background lights; and (v) difficulties in detection of LEDs by PDs, especially under direct sunlight. Furthermore, with increasing transmission distance, small-area PDs yield reduced optical receiving power, thus limiting the transmission range. These limitations can be easily overcome using a camera/IS as the receiver.

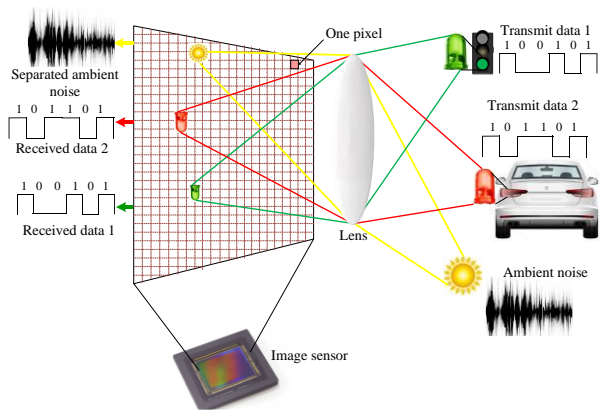

Fig. 8. Operating principle of OCC system.

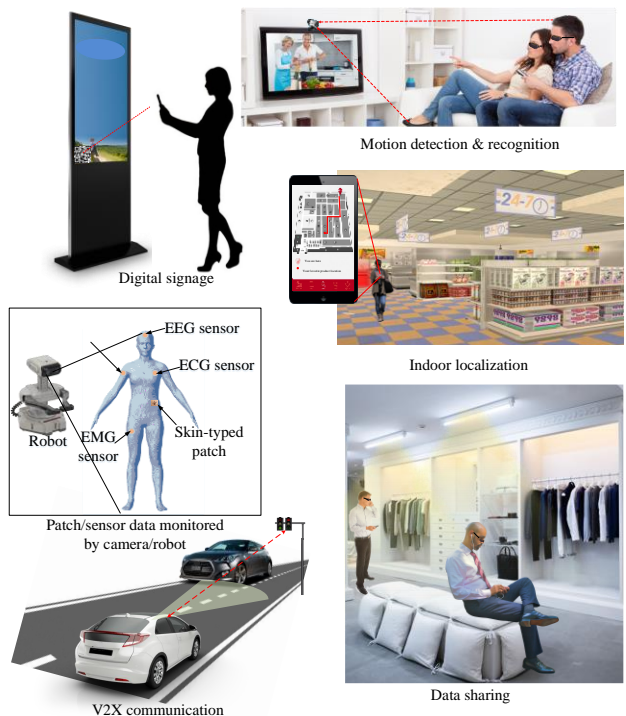

Fig. 9. Application scenarios of OCC system.

To increase the level of received optical power, a concentrator is usually employed in front of a PD [21]. Although a VLC system employing a concentrator with small-area PD can provide a data rate of the order of several Gbps, the limitations that cannot be overcome are shorter range, difficult in separating each of the multiple optical signals, and lack of mobility. Therefore, currently, these factors limit the practical application of VLC, especially for comparatively long-range communication and outdoor V2X communication.

To overcome the limitations of PDs as receivers in LED/PD-based VLC systems, several works on IS-based optical signal reception have already been reported [90]–[107]. In [105], CMOS technology was used to develop an OWC system equipped with an optical communication image sensor (OCI), and the design and fabrication of the OCI, as well as the development of an IS-based OWC system with a 20 Mbs/pixel data rate without LED detection and a 15 Mbs/pixel data rate with a 16.6 ms real-time LED detection, were described. In

[34], speeds of 45 Mbps without bit errors and 55 Mbps with a bit error rate of $10^{-5}$ were demonstrated using an OCC system.

An optical IS or camera composed of a PD array is used as a receiver in OCC. By using an image lens in front of IS, light from various sources with different directions within the field of view (FOV) of the camera is projected onto various positions on an IS and is sampled by different pixels [21]. A camera can be modeled as a two-dimensional PD array, which usually contains more than 10 million pixels and each pixel measures 1–10 μm in size [108], [109]. Therefore, a camera has a very high spatial resolution, which means that light coming from slightly different directions can be projected onto various positions on a sensor by using a lens and be sampled by different pixels. This feature enables it to separate lights from different sources and different directions, which is ideal for spatial-division multiplexing and imaging MIMO systems in VLC [21]. Fig. 8 shows the basic principle of an OCC system. As shown in the figure, the data transmitted from two different LED transmitters (vehicle rear LED array and traffic light) can be captured easily and distinguished simultaneously by using IS. Background noise sources (e.g., sunlight, ambient light, and digital signage) can be discarded by separating the pixels associated with such noise sources. In this manner, IS can provide secure, interference-free, and reliable communications even in outdoor conditions.

Fig. 9 show a few application scenarios of OCC systems. OCC system can be used for many indoor and outdoor applications. Viewers can receive information from digital signage by using their smartphone cameras. Data monitored by wearable skin patches can be collected by using any camera. Very precise cm-level indoor and outdoor localization can be performed using OCC technology. OCC is very promising for vehicle-to-vehicle, vehicle-to-infrastructure, and infrastructure-to-vehicle communications. Mobile users can share their data though OCC technology. OCC can be used to perform motion detection and recognition. User can send control signals by using body motion, and an OCC system can be used to detect such signals. A few important advantages and limitations of OCC systems are listed below:

**Main advantages of OCC:**

- Non-interference communication: When an IS is used as a receiver, lights from different sources are separated almost perfectly on a focal plane because there is a huge number of pixels in IS, and optical signals are output separately from each pixel, which avoids signals from becoming mixed, thus permitting communication even if light signals and ambient lights such as sunlight and streetlights are present [34], [105].
- High SNR quality: OCC system provides very high SNR level and therefore requires no complex protocol for simultaneous communication among multiple LEDs.
- No need for complex signal processing: The LEDs that are not required by receivers can be omitted entirely from IS because users can easily select communication partners from an image, and therefore, there is no need for complex signal processing to filter unnecessary information [105].

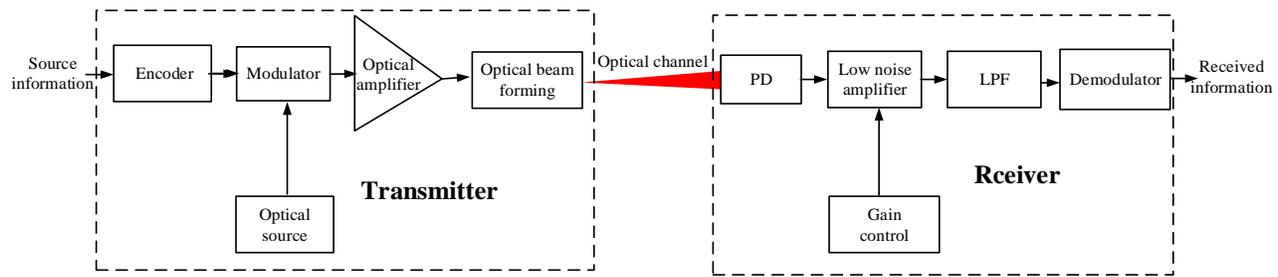

Fig. 10. Basic block diagram of an FSO system [37].

- Stable against changing communication distances: The optical signal power received by an IS is stable against a changing communication distance [105]. LED image size is decreased with an increase in communication distance. The incident light power per pixel remains unaffected in spite of variations in communication distance as long as the imaged LED size on the focal plane of the IS is not smaller than the pixel size [106]. Therefore, a one-pixel size LED image is sufficient to perform stable communication between the LED and IS. Hence, stable communication can be achieved using an OCC system as long as the imaged LED size is not smaller than the size of a pixel.

**Main limitations of OCC:**

- Blocking of optical link: An optical channel is LOS, and therefore, its communication links will be blocked by any object that prevents light penetration, such as walls, buildings, thick fog, and thick gas [105].
- Low-data-rate: Currently, OCC cannot support high-data-rate communication owing to the low frame rates of conventional cameras. The frame rate of commonly used ISs is 30 fps, which is not adequate for achieving high-data-rate communication. The data rate in terms of bps/pixel must be limited to half of the rate fps or less to satisfy the Nyquist frequency requirement. Therefore, to attain higher data rates, high-speed cameras or newly developed ISs should be used in the receiver systems [105].
- Light flickering: Owing to the typical low frame rates of commercial cameras, if a signal is transmitted at a low frequency by using VL sources, the human eye will detect light flickering, which is not acceptable for light illumination [107].

OCC has a wide range of applications. A few important applications are listed below:

- V2X communication: Cameras are being used in vehicles. Through combining VL/NIR-based OWC technology and these cameras, OCC systems that provide the functionalities of monitoring and data communications simultaneously can be developed. Therefore, OCC technology is very attractive for autonomous vehicle.
- Indoor positioning: One of the potential applications of OCC systems is high-accuracy indoor positioning. If each LED used in an indoor environment is given a unique identification (ID) code, a smart device with a built-in camera can be used to effectively locate devices, people, and objects within a room [21].

- Digital signage: Nowadays digital signage or display has become the most popular medium among businesses for broadcasting advertisements or messages and offering coupons to customers. Viewers can use their smartphones to receive information from digital signage. This will provide real-time screen to camera communication behind any scene.
- Drone-to-drone communication: In the next generation, the world will be monitored using drones. Therefore, communication among drones is necessary to avoid drone-to-drone collisions. OCC can be a future candidate for drone-to-drone communication.
- Augmented/virtual reality: Virtual or augmented realization is an upcoming technology. OCC can support these augmented and virtual reality services and provide an improved approach by adapting certain machine-learning approaches (e.g., artificial neural networks and convolution neural networks).

### D. Free Space Optical Communication (FSOC)

FSOC [37]–[39], [110]–[125], a subset of OWC, is normally operated using the NIR spectrum as the communication medium. It can be also operated by using the VL and the UV spectra. However, when using IR, the attenuation levels are lower. Illumination is not required in FSO. FSO often uses LDs rather than LEDs for the transmission. Narrow beams of focused light from an LD transmitter are used to establish high-data-rate communication links between a transmitter and a receiver. FSO systems are used for high-data-rate communication between two fixed points over distances ranging from a few nm to several thousand kilometers [37]. FSO systems use laser technology for signal transmission. Fig. 10 shows the basic block diagram of an FSO system. Because of optical beamforming, long-distance communication is possible using FSO systems. The source information data input is to be transmitted to a long destination. Initially, the source information is encoded. Optionally, channel coding can be used before modulation. The modulator performs an OOK/PM/OFDM modulation process. If required, an optical amplifier can be used to increase the power intensity of the modulated laser beam. Then, the light beam is collected and refocused by means of beam forming optics before being transmitted [37]. The typical optical source in FSO systems is an LD [52]. A few manufacturers also use high-power LEDs with beam collimators. The optical source used in FSO systems should deliver a comparatively high optical power over a wide temperature range. The important features of the optical transmitter in FSO systems are size, power, and beam quality, which determine the laser intensity and minimum divergence obtainable from the system [37]. The receiver front-end comprises optical filters with a lens to collect and focus the received beam onto the PD. The PD output current is next converted to voltage. The low-pass filter is used to limit the thermal and background noise levels. Finally, the demodulator performs the necessary demodulation process to obtain the original sending data.

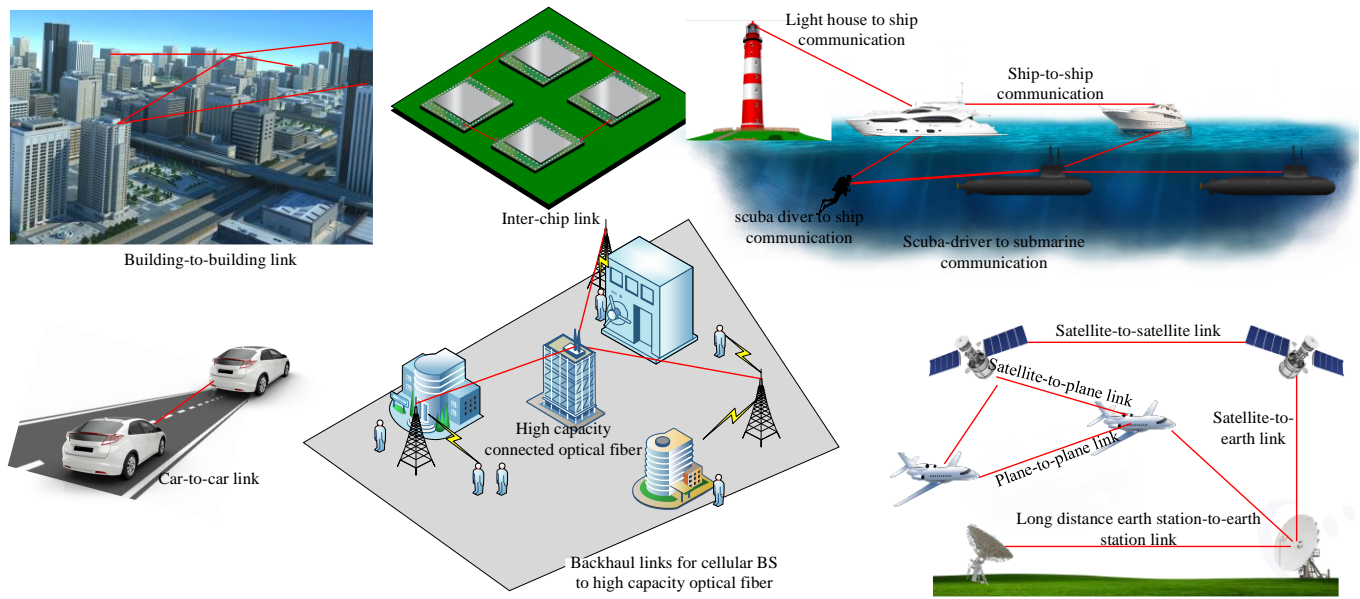

Fig. 11. A few application scenarios of FSO systems.

The frequency used for FSOC is higher than 300 GHz, which is totally unregulated worldwide. In comparison to RF-based communication, FSO links have very high available bandwidth, thus providing considerably higher data rates. Transmission rates of 10 Gbps have already been implemented for long-distance communication [125]. A 40-Gbps FSO link was implemented in [38] over a communication distance of 20 m. The data rates of recent FSOC systems are comparable to those achieved using fiber optics [125]–[129]. FSOC systems can be deployed easily. Therefore, they have a wide range of applications [37], [38], [120], [130]–[133], including uses such as backhaul for cellular networks, disaster recovery, high-definition TV, medical image/video transmission, campus connectivity, video surveillance and monitoring, security, and broadcasting. Other applications of FSOC systems include inter-chip connectivity, MAN-to-MAN connectivity, LAN-to-LAN connectivity, ship-to-ship connectivity, underwater communication, airplane-airplane/ground/satellite connectivity, satellite-satellite/earth/airplane connectivity, and fiber back-up. Fig. 11 shows a few application scenarios of FSO systems. In this figure, nm-range optical interconnection, car-to-car connectivity, links among different buildings, cellular backhaul connectivity, underwater communication, ship-to-ship or ship-to-infrastructure connectivity, satellite-to-satellite/earth/airplane connectivity, and airplane-to-airplane/earth connectivity using FSOC are shown. As can be seen, FSOC can cover a very wide range of wireless connectivity types.

TABLE V
DIFFERENCE BETWEEN FSO COMMUNICATION AND MICROWAVE LINK

| Issue | FSO Communication | Microwave Link |
|---|---|---|
| Transmission medium | NIR, VL, or UV | Millimeter waves |
| Maximum communication distance | More than 10,000 km | More than 100 km |
| Data rate | 40 Gbps [38] at a communication distance of 20 m, 5.6 Gbps with LEO-LEO [125] | 12.5 Gbps at communication distance of 5.8 m [134] |
| Interference | Low | High |
| Environmental impact | High | Low |
| Infrastructure cost | Less | High |

Despite the many advantages of FSOC across a wide range of applications, it suffers from link reliability, especially in long-range communications, owing to its high sensitivity to several factors. These include weather conditions (e.g., fog, rain, haze, smoke, sandstorms, clouds, snow, and scintillation can increase attenuation), atmospheric turbulence (causing fluctuations in the density of air, leading to a change in the air refractive index), and physical obstructions (e.g., flying birds, trees, and tall buildings can temporarily block the signal beam) [40]. Table V shows some important differences between FSOC and microwave links.

### E. Light Detection and Ranging (LiDAR)

LiDAR is an attractive optical remote sensing technology that finds the range of and/or other information about a distant target [42], [43], [135]–[141]. LiDAR uses normally NIR and VL to image objects. It can target a wide range of materials including rain, dust, non-metallic objects, chemical compounds, aerosols, clouds, and even individual molecules [135]. A narrow laser beam can map physical features with very

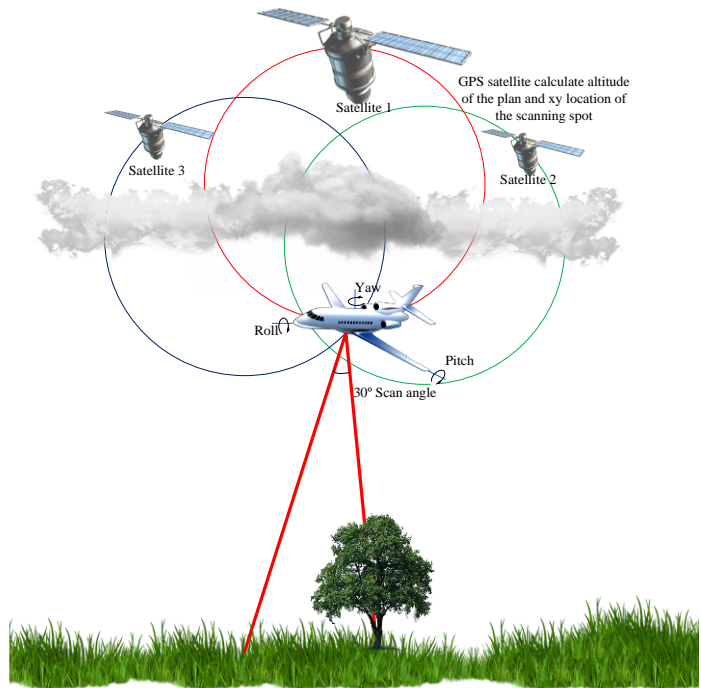

Fig. 12. LiDAR application in airplane

TABLE VI

DIFFERENCES BETWEEN LiDAR AND RADAR

| LiDAR | RADAR |
|---|---|
| It uses laser light rays (NIR, visible) for transmission and reception medium. | It uses radio waves ( microwave signals in the range of 1 cm wavelength) for transmission and reception medium |
| It can provide accurate surface measurements with 3D mapping | Size and the position of the object can be identified fairly using RADAR |
| It uses charge coupled device (CCD) optics and lasers for transmission and reception | It uses antennae for transmission and reception of the signals |
| Lower wavelengths allow detection of very small objects e.g., cloud particles | Target size is limited by longer wavelength |
| Performance degraded with bad atmospheric condition | It can operate in bad weather conditions |

high resolution, for example, a laser mounted on an aircraft can be used to map terrain at a resolution of 30 cm or better [136]. Laser beams are used to measure the properties of scattered light and create points for 3D mapping. The four main components of a LiDAR system are laser, photodetector and receiver electronics, scanner and optics, and position and navigation sub-systems. The purposes of LiDAR are similar to those of radio detection and ranging (RADAR). Both LiDAR and RADAR use similar technologies and concepts to track the position and the movements of objects. There are differences in how each technology works and the types of applications for which each can be best used. Both technologies use energy reflected from target objects to determine various aspects of those objects. However, the type of energy used by LiDAR is optical light, whereas RADAR uses microwaves. In both LiDAR and RADAR, energy is transmitted from a transmitter as a signal. When the transmitted signal hits an object, the object reflects a part of the energy of the original signal. This reflected energy is then received by the receiver at the source location and is used to determine the distance, size, and other characteristics of the object. The types of objects that can be located precisely and measured using RADAR and LiDAR are also different in size and nature. Table VI shows the main differences between LiDAR and technologies.

Three platforms are used to collect LiDAR data: ground, airplane, and space. Fig. 12 shows an example of airborne LiDAR, which is most commonly used platform. The principle of airborne LiDAR data collection is divided into four steps. First, a side-by-side laser scan of the ground is performed. Second, the altitude and the x-y location of the airplane are obtained using GPS. Third, an inertial measurement unit is used to track the tilt of the plane in the sky as it flies for accurate calculation of alleviation. Finally, all data are accumulated and range-related information is then calculated [138]. Fig. 13 shows the application of LiDAR data from ground, which is mostly used in autonomous vehicles. It calculates the distance between a car and an object and builds a 3D image of its environment. LiDAR data is combined with radar sensor information and processed by the onboard computing system, which then makes decisions about steering, breaking, and acceleration [139]. Fig. 13 (a) shows a real scenario from a vehicle, and Fig. 13 (b) shows a 3D map of objects constructed using LiDAR for the same scenario.

A few useful applications of LiDAR include meteorology, autonomous vehicles, transportation, architectural surveys, military, space exploration, robot vision, precision guidance, and vehicle anti-collision. The LiDAR used for localization and obstacle detection in [135] delivers 1.6 million 3D points per second. A few applications of LiDAR are listed below:

- Transportation: LiDAR images can show road contour, lane detection, localization, elevation, and roadside vegetation for safe transportation.
- Airborne LiDAR: It is a laser scanner attached to a plane. It creates a 3D point cloud model of the landscape during flight.
- Autonomous vehicles: Autonomous vehicles use LiDAR for obstacle detection and avoidance to navigate securely through environments by using rotating laser beams.
- Forestry: LiDAR has many important applications in forestry. Canopy heights, biomass measurements, and leaf area can be studied using airborne LiDAR systems.
- Geology and soil science: High-resolution digital elevation maps created using LiDAR technology have led to significant developments in geomorphology.
- Physics and astronomy: A worldwide network of observatories uses LiDARs to measure the distance to reflectors placed on the moon, allowing for measurement of the position of the moon with mm precision and tests of general relativity to be conducted.

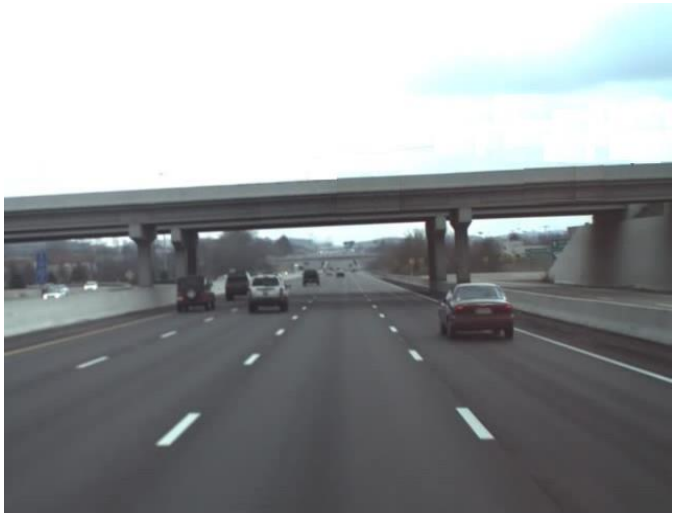

(a)

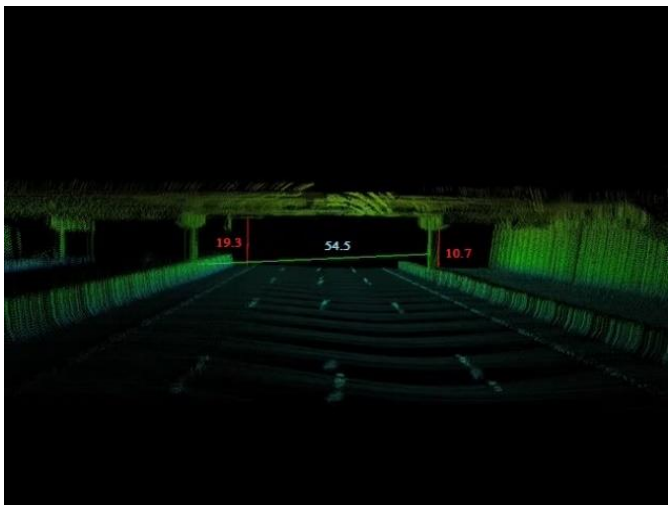

(b)

Fig. 13. LiDAR application in vehicle [140], (a) real environment, (b) 3D image mapped by LiDAR.

- Archaeology: In the field of archaeology, LiDAR has some important applications including supporting the planning of field campaigns, mapping features beneath forest canopy, and providing overviews of broad, continuous features that may be indistinguishable on the ground. In addition, LiDAR provides archaeologists the capability to make high-resolution digital elevation models of archaeological sites.
- Military: High-resolution LiDAR systems gather sufficient information to identify targets such as tanks. Military applications of LiDAR include the Airborne Laser Mine Detection System for counter-mine warfare.
- Rock Mechanics: LiDAR has been used widely in the area of rock mechanics for rock mass characterization and slope change detection. A few important geomechanical

- properties such as discontinuity orientation, discontinuity aperture, and discontinuity persistence of rock mass can be extracted from the 3D point clouds obtained using LiDAR [141].
- Robotics: LiDAR is used in robotics for observing the environment and for object classification [142]. LiDAR can provide 3D elevation maps of a territory with very high precision distance to the ground. This approach can help calculate safe landing velocities of robotic and manned vehicles with a high degree of accuracy.

## IV. SUMMARY, OPEN ISSUES, AND FUTURE RESEARCH DIRECTIONS

### A. Summary

Different wireless communication technologies have different characteristics. They follow different standards, use modulation techniques, transmission media, physical transmitters, physical receivers, have different principles of operation, attenuation characteristics, and security level. Therefore, based on our review of different OWC technologies, Table VII summarizes a comparison among different wireless communication systems. This table clearly shows the differences in the basic principles, architectures, and application scenarios among the wireless technologies. It also shows the clear advantages and disadvantages of OWC technologies. VLC can be used for illumination, communication, and localization whereas LiFi can be used for illumination and communication. Both VLC and LiFi can provide high-data-rate communication, but they are not very effective for outdoor applications due to interference from natural light and other light sources. Moreover, they cannot provide long-distance communication. OCC can be used for communication, imaging, and positioning purposes. It can provide good performance in terms of communication over comparatively long distances, even in outdoor applications. The main drawback of OCC is its low data rate due to the low frame rates of conventional cameras. FSOC can support high-data-rate and ultra-long-range communications, but its performance is affected by the environment. LiDAR is an excellent sensing technology for very-high-resolution 3D mapping. However, currently, this system is very expensive and its performance is affected by the environment. Therefore, the areas of application of each of the technologies should be different for the effective use of them. The maximum data rate of 100 Gbps is found in OWC whereas maximum 6 Gbps is found in RF-based communication. It is observed that the communication distance using RF-based technology is lower than that of OWC technology. The interference effect on OWC is also lower than that of RF-based communication. Moreover, the level of security in OWC is higher than RF-based communication. However, OWC systems suffer owing to their sensitivity to blocking by obstacles. Therefore, the coexistence of OWC and RF systems is an effective solution for future wireless commutations.

TABLE VII
SUMMARY OF COMPARISON OF DIFFERENT WIRELESS COMMUNICATION SYSTEMS.

| Issue | LiFi | VLC | OCC | FSO | RF |
|---|---|---|---|---|---|
| Standardization | In progress (previously it was IEEE 802.15.7m TG Task Group (TG) and now changed to IEEE 802.15.11 LC SG) | Matured (IEEE 802.15.7-2011) | In progress (IEEE 802.15.7m TG) | Well developed | Matured |
| Transmitter | LED/LD(combined LDs with optical diffuser) | LED/LD | LED | LD | Antenna |
| Receiver | PD | PD/camera | Camera | PD | Antenna |
| Modulation | Amplitude and phase modulation techniques cannot be applied. Information need to be modulated in the varying of intensity of the light wave. On-Off Keying (OOK), Pulse modulation (PM), Orthogonal Frequency Division Modulation (OFDM), Code-Division Multiple-Access (CDMA), Color Shift Modulation (CSK) etc. are applicable [26], [143], [144]. | Amplitude and phase modulation techniques cannot be applied. Information need to be modulated in the varying of intensity of the light wave. OOK, PM, CDMA, OFDM, CSK etc. are applicable [26], [143], [144]. | Amplitude and phase modulation techniques cannot be applied. Information need to be modulated in the varying of intensity of the light wave. OOK, PM, CDMA, OFDM, CSK etc. are applicable [26], [143], [144]. In addition, screen-based modulation [99] is also applicable for OCC. | Amplitude and phase modulation techniques cannot be applied. Information need to be modulated in the varying of intensity of the light wave. OOK, PM, OFDM, etc. are applicable [26], [143], [145]. | ASK, PSK, FSK, PM, CDMA, OFDM, etc. are applicable. |
| OFDM | Yes [26] | Yes [26] | Yes [34] | Yes [143] | Yes |
| MIMO | Yes [80] | Yes [21] | Yes [105] | Yes [110] | Yes |
| Communication distance | 10 m [81] | 20 m | 200 m | More than 10000 km | More than 100 km using Microwave link |
| Interference level | Low [15] | Low [26] | Zero [21] | Low [111] | Very high |
| Noise | Sun plus ambient light sources | Sun plus ambient light sources | Sun plus ambient light sources | Sun plus ambient light sources | All electrical and electronic appliances |
| Environmental effect | Indoor: No Outdoor: Yes | Indoor: No Outdoor: Yes | No | Yes | Yes |
| Data rate | 10 Gbps using LED and 100 Gbps using LD [12] | 10 Gbps using LED and 100 Gbps using LD [12] | 54 Mbps [34] | 40 Gbps [37] | 6 Gbps [146] (IEE 802.11ad at frequencies around 60 GHz) |
| Security | High | High | High | High | Low |
| Spectrum | IR/VL/UV | VL | IR/VL | IR/VL/UV | Radio waves |
| Spectrum regulation | No | No | No | No | Yes (not always e.g., WiFi) |
| Path loss | Medium ( very high for NLOS) | Medium ( very high for NLOS) | Less | High | High |
| Illumination | Yes | Yes (only when LED bulb is used) | No | No | No |
| Main purpose | Illumination and Communication | Communication, illumination, and localization | Communication, imaging, and localization | Communication | Communication and positioning |
| Main limitations | (i) Short distance communication, and (iii)  not suitable in outdoor | (i) Short distance communication, (ii) no guaranteed of mobility support, and (iii) not suitable in outdoor | Low data rate | Environment dependent | Interference |

## B. Open Issues and Future Research Directions

A few important research issues and future directions of OWC technologies are briefly discussed below:

**Hybrid network architecture**: This approach integrates two or more different technologies (e.g., OWC/RF, FSO/RF, WiFi/LiFi, VLC/femtocell, VLC/FSO, and LiFi/OCC) into a hybrid network [82], [147] and capable of providing the advantages of both the technologies. Especially, optical wireless is a good candidate for sensitive applications in which mitigating interference with RF is a must. Hybrid network can play an important role for load balancing, link reliability improvement, wireless connectivity availability in remote places (e.g., deep-space, deep-ocean, and deep-ground situations), and interference reduction. Therefore, this issue attracted the researchers. Very important challenge of hybrid system is to adaptively and smoothly switch from one communication system to another.

**High-data-rate optical backhaul:** For 5G and beyond communications, a high capacity backhaul connectivity is very important issue [3]. The optical wireless network e.g., FSO or VLC can be a good complementary choice of the existing wired and wireless backhaul connectivity, and is a promising research issue.

**NLOS UV communication:** A high-data-rate NLOS communication is achievable using UV communication that is a significant feature of UV band [23]. A methodical assessment and exploitation of this communication in terms of atmospheric conditions, geometrical configurations, and transmitter/receiver system requirements is still missing and should be the issue of future research.

**Inter-cell interference:** The dense deployment of LEDs for VLC/LiFi small cell/attocell architecture faces a challenge of managing inter-cell optical interference [148]. Therefore, researchers are focusing on this issue.

**Extending the optical spectrum:** Extending the optical spectrum beyond the UV band can enjoy a lot of benefits from high power and low cost sources [15]. Thus, this is a challenging issue for future wireless communication.

**Underwater wireless optical communication:** In recent years, underwater wireless optical communication attracted much attention as many applications are proposed for environmental monitoring, oil pipe investigation, and offshore investigation. Long-range and high-speed links are needed for many applications of underwater communication. An employing of 405 nm blue light LD is expected to be an important research issue for a long-range underwater communication. The design of appropriate modulation and coding techniques that can adapt the characteristics of underwater environments is also an important research direction [149].

**Seamless mobility:** OWC systems are required to allow the user mobility. Currently, only LiFi provides seamless connectivity [31]. Therefore, the researchers are working to support horizontal handovers in OWC (e.g., between LiFi networks) [31] as well as vertical handovers [44] in hybrid network (e.g. between LiFi and WiFi networks). It is very important to have handover mechanism to maintain seamless communications, and such an issue is challenging and need to be properly studied.

**Atmospheric loss:** In OWC, the power allocation and overall system performance of a free space propagation link are strongly dependent on atmospheric loss which comprises scattering, refraction, clear air absorption, free space loss, and scintillation [40]. Hence, there is huge scope to work for the mitigation of the atmospheric loss in OWC systems.

**MIMO optical wireless communications:** MIMO is already a well-established technique in RF-based communication systems. However, because of the properties of the IM/DD channel, it is challenging to apply MIMO in OWC.

The applicability of MIMO for different OWC technologies are under investigation by many researchers [21], [80], [105], [110]. The MIMO system benefits from higher power efficiency, higher reliability, and higher capacity. However, additional device introduced by MIMO increases the complexity and also limits its applications (e.g., several compact sensors underwater communication). Moreover, the narrow beamwidth of LED receivers have limited angles of views, and thus small misalignment between a transmitter and a receiver can easily disrupt the communication. Therefore, MIMO system with considering accurate channel model for different OWC technologies is a potential research issue.

**OWC in health application:** Currently, various RF-based technologies are used in healthcare applications. Several medical devices are very sensitive to electromagnetic interference by RF [150] and can critically affect the performance of many medical devices. Moreover, RF radiation is not good for humans. LED-based OWC can offer an excellent solution for the raised complications due to the usage of RF-based wireless technologies in healthcare and should be aim of future research.

**OWC in vehicular communication:** A standard IEEE 802.11p which is also known as Wireless Access in Vehicular Environments (WAVE), uses unlicensed RF band [151]. OWC system can be also applied for traffic management [152]. Hence, vehicular communication based on OWC technologies can be a good research issue.

**OWC for positioning:** The RF-based indoor positioning systems are less accurate mostly due to multipath induced fading and signal penetration. The positioning using LEDs have attracted much attention due to the high accuracy, license-free operation, no electromagnetic interference, and low cost frontends, etc. [101]. Outdoor positioning using optical wireless technology can also be one potential future research issue.

**OWC in drone application:** Drone is thinking as the next possible solution within transportation [153]. Its large scale deployment will arise several issues e.g., precise localization and reliable communication among drones. OWC can be a good complementary approach to solve these issues due to the limitations of RF-based technologies.

**Data rate improvement of OCC system**: Currently, the achieved data rate of OCC system is not high (55 Mbps until now [34]). This data rate should be increased to fulfill the upcoming demands of high-data-rate services.

**OCC based on red, green, and blue (RGB) LEDs:** OCC using RGB LEDs and color cameras (or IS) is a promising approach for effective parallel visible light communications [50], [84], [145]. Many researchers are working on RGB LEDs for developing efficient communication technology.

**Dimming control:** The algorithms for dimming [79] should be designed for the effective deployment of OWC systems.

**Flickering avoidance:** Flickering is the fluctuations in the brightness of a light that can be noticed by humans. It is a harmful effect on human health [79]. In OWC systems, the modulation of the LEDs should be done in a way that the flickering is avoided. Therefore, this is an important research issue for the deployment of OWC systems.

## V.  CONCLUSIONS

The current RF system cannot fulfill the exponential growth in demand of wireless capacity for 5G and beyond communications. The extremely wide optical band that includes infrared, visible, and ultraviolet sub-bands, can be used for wireless connectivity to support 5G and IoT. The use of optical bands complementarily to RF will relieve the problems caused by spectrum shortage in RF-based wireless communication. Researchers are trying to adapt many applications that are currently based on RF communication to OWC. Optical wireless technologies will play very important roles in industry, healthcare, public gathering places, stadiums, transportation, residences, offices, shopping mall, underwater communication, space. In this survey paper, we have presented a general overview of emerging OWC technologies. We categorized OWC systems from different viewpoints. The main optical wireless technologies discussed in this paper include VLC, LiFi, OCC, FSOC, and LiDAR. These can be used for ultra-short distance to ultra-long distance communication. Each of these technologies has unique advantages and limitations. Therefore, the areas of application of each of the technologies are different. Accordingly, we have presented clear differences in the basic principles, architectures, and application scenarios considering all of the mentioned optical wireless technologies. In addition, we have explained the challenges associated with each of the technologies. Indeed, optical wireless technologies are promising for securing a better life in the future. We hope that this comparative survey will serve as a valuable resource for understanding the research contributions in the emerging optical wireless technologies and hopefully prompt further efforts for the successful deployment of OWC systems as a prominent complementary to RF-based technologies in the future 5G and beyond heterogeneous wireless networks.

## REFERENCES

[1] P. Pirinen, "A brief overview of 5G research activities," in Proc. of *International Conference on 5G for Ubiquitous Connectivity (5GU)*, Nov. 2014, pp. 17-22.

[2] M. Shafi, A. F. Molisch, P. J. Smith, T. Haustein, P. Zhu, P. D. Silva, F. Tufvesson, A. Benjebbour, and G. Wunder, "5G: a tutorial overview of standards, trials, challenges, deployment, and practice," *IEEE Journal on Selected Areas in Communications*, vol. 35, no. 6, pp. 1201-1221, Jun. 2017.

[3] M. Jaber, M. A. Imran, R. Tafazolli, and A. Tukmanov, "5G backhaul challenges and emerging research directions: a survey," *IEEE Access*, vol. 4, pp. 1143-1166, May 2016.

[4] D. Zhang, Z. Zhou, S. Mumtaz, J. Rodriguez, and T. Sato, "One integrated energy efficiency proposal for 5G IoT communications," *IEEE Internet of Things Journal*, vol. 3, no. 6, pp. 1346-1354, Dec. 2016.

[5] H. A. U. Mustafa, M. A. Imran, M. Z. Shakir, A. Imran, and R. Tafazolli, "Separation framework: an enabler for cooperative and d2d communication for future 5G networks," *IEEE Communication Surveys & Tutorials*, vol. 18, no. 1, pp. 419-445, 2016.

[6] J. G. Andrews, S. Buzzi, W. Choi, S. V. Hanly, A.Lozano, A. C. K. Soong, and J. C. Zhang, "What will 5G be?," *IEEE Journal on Selected Areas in Communications*, vol. 32, no. 6, pp. 1065-1082, Jun. 2014.

[7] A. A. Fuqaha, M. Guizani, M. Mohammadi, M. Aledhari, and Moussa Ayyash, "Internet of things : a survey on enabling technologies, protocols, and applications," *IEEE Communication Surveys & Tutorials*, vol. 17, no. 4, pp. 2347-2376, 2015.

[8] P. Schulz, M. Matthé, H. Klessig, M. Simsek, G. Fettweis, J. Ansari, S. A. Ashraf, B. Almeroth, J. Voigt, I. Riedel, A. Puschmann, A. M. Thiel, M. Müller, T. Elste, and M. Windisch, "Latency critical IoT applications in 5G: perspective on the design of radio interface and network architecture," *IEEE Communications Magazine*, pp. 70-78, Feb. 2017.

[9] M. R. Palattella, M. Dohler, A. Grieco, G. Rizzo, J. Torsner, T. Engel, and Latif Ladid, "Internet of things in the 5G era: enablers, architecture, and business models," *IEEE Journal on Selected Areas in Communications*, vol. 34, no. 3, pp. 510-527, March 2016.

[10] W. A. Hassan, H.-S. Jo, and T. A. Rahman, "The feasibility of coexistence between 5G and existing services in the IMT-2020 candidate bands in Malaysia," *IEEE Access*, vol. PP, no. 99, 2017.

[11] A. Ijaz, L. Zhang, M. Grau, A. Mohamed, S. Vural, A. U. Quddus, M. A. Imran, C. H. Foh, and R. Tafazolli, "Enabling massive IoT in 5G and beyond systems: phy radio frame design considerations," *IEEE Access*, vol. 4, pp. 3322-3339, Jul. 2016.

[12] D. Tsonev, S. Videv, and H. Haas, "Towards a 100 Gb/s visible light wireless access network," *Optics Express*, vol. 23, no. 2, pp. 1627-1637, Jan. 2015.

[13] L. Hanzo, H. Haas, S. Imre, D. O'Brien, M. Rupp, and L. Gyongyosi, "Wireless myths, realities and futures: from 3G/4G to optical and quantum wireless," *Proceedings of IEEE*, vol. 100, pp. 1853-1888, May 2012.

[14] M. Uysal and H. Nouri, "Optical wireless communications – an emerging technology," in Proc. of *International Conference on Transparent Optical Networks*, Jul. 2014.

[15] Z. Ghassemlooy, S. Arnon, M. Uysal, Z. Xu, and J. Cheng, "Emerging optical wireless communications-advances and challenges," *IEEE Journal on Selected Areas in Communications*, vol. 33, no. 9, pp. 1738-1749, Sep. 2015.

[16] A. Boucouvalas, P. Chatzimisios, Z. Ghassemlooy, M. Uysal, and K. Yiannopoulos, "Standards for indoor optical wireless communications," *IEEE Communication Magazine*, vol. 53, no. 3, pp. 24-31, March 2015.

[17] H. Chen, H. P. A. Boom, E. Tangdiongga, and T. Koonen, "30-Gb/s bidirectional transparent optical transmission with an MMF access and an indoor optical wireless link," *IEEE Photonics Technology Letters*, vol. 24, no. 7, pp. 572-574, April 2012.

[18] D. K. Borah, A. C. Boucouvalas, C. C. Davis, S. Hranilovic, and K. Yiannopoulos, "A review of communication-oriented optical wireless systems," *EURASIP Journal on Wireless Communications and Networking*, vol. 1, no. 1, pp. 1-28, March 2012.

[19] S. Dimitrov and H. Haas, "Information rate of OFDM-based optical wireless communication systems with nonlinear distortion" *Journal of Lightwave Technology*, vol. 31, no. 6, pp. 918-929, March 15, 2013.

[20] H. Elgala, R. Mesleh, and H. Haas, ''Indoor optical wireless communication: Potential and state-of-the-art,'' *IEEE Communication Magazine*, vol. 49, no. 9, pp. 56-62, Sep. 2011.

[21] Z. Ghassemlooy, P. Luo, and S. Zvanovec, *Optical Camera Communications*, pp 547-568, Springer, Aug. 2016.

[22] J. B. Carruthers, "Wireless infrared communications," *Wiley Encyclopedia of Telecommunications*, 2003.

[23] Z. Xu and R. Br. M. Sadler, "Ultraviolet communications: potential and state-of-the-art," *IEEE Communications Magazine*, vol. 46, no. 5, pp. 67-73, May 2008.

[24] Z. Xu, G. Chen, F. A.-Galala, and M. Leonardi, "Experimental performance evaluation of non-line-of-sight ultraviolet communication systems," in Proc. of *SPIE Photonics and Optics*, Aug. 2007, pp. 1-12.

[25] H. Qin, Y. Zuo, D. Zhang, Y. Li, and J. Wu, "Received response based heuristic LDPC code for short-range non-line-of-sight ultraviolet communication," *Optics Express*, vol. 25, no. 5, pp. 5018-5030, March 2017.

[26] P. H. Pathak, X. Feng, P. Hu, and P. Mohapatra, "Visible light communication, networking, and sensing: a survey, potential and challenges," *IEEE Communications Surveys & Tutorials*, vol. 17, no. 4, pp. 2047-2077, 2015.

[27] F. Yang and J. Gao, "Dimming control scheme with high power and spectrum efficiency for visible light communications," *IEEE Photonics Journal*, vol. 9, no. 1, pp. 1-12, Feb. 2017.

[28] Y.-J. Zhu, Z.-G. Sun, J.-K. Zhang, Y.-Y. Zhang, and J. Zhang, "Training receivers for repetition-coded MISO outdoor visible light communications," *IEEE Transactions on Vehicular Technology*, vol. 66, no. 1, pp. 529-540, Jan. 2017.

[29] S. Dimitrov and H. Haas, *Principles of LED Light Communications: Towards Networked Li-Fi.* Cambridge, U.K.: Cambridge University Press, March 2015.

[30] H.-H. Lu, C.-Y. Li, H.-W. Chen, C.-M. Ho, M.-T. Cheng, Z.-Y. Yang, and C.-Kai Lu, "A 56 Gb/s PAM4 VCSEL-based LiFi transmission with two-stage injection-locked technique," *IEEE Photonics Journal*, vol. 9, no. 1, pp. 1-8, Feb. 2017.

[31] H. Haas, L. Yin, Yunlu Wang, and Cheng Chen, "What is LiFi?," *Journal of Lightwave Technology*, vol. 34, no. 6, pp. 1533-1544, March 15, 2016.

[32] Y.-C. Chi, D.-H. Hsieh, C.-Y. Lin, H.-Yu Chen, C.-Y. Huang, J.-H. He, B. Ooi, S. P. Den Baars, S. Nakamura, H.-Chung Kuo, and Gong-Ru Lin, "Phosphorous Diffuser Diverged Blue Laser Diode for Indoor Lighting and Communication," *Nature Scientific Reports*, Dec. 2015.

[33] D. Tsonev, H. Chun, S. Rajbhandari, J. J. D. McKendry, S. Videv, E. Gu, M. Haji, S. Watson, A. E. Kelly, G. Faulkner, M. D. Dawson, H. Haas, and D. O'Brien, "A 3-Gb/s single-LED OFDM-based wireless VLC link using a gallium nitride μLED," *IEEE Photonics Technology Letters*, vol. 26, no. 7, pp. 637-640, April 2014.

[34] Y. Goto, I. Takai, T. Yamazato, H. Okada, T. Fujii, S. Kawahito, S. Arai, T. Yendo, and K. Kamakura, "A new automotive VLC system using optical communication image sensor," *IEEE Photonics Journal*, vol. 8, no. 3, pp.1-17, Jun. 2016.

[35] T. Yamazato et al., "Image-sensor-based visible light communication for automotive applications," *IEEE Communications Magazine*, vol. 52, no. 7, pp. 88-97, Jul. 2014.

[36] Z. Zheng, L. Liu, and W. Hu, "Accuracy of ranging based on DMT visible light communication for indoor positioning," *IEEE Photonics Technology Letters*, vol. 29, no. 8, pp. 679-682, April 2017.

[37] M. Ali Khalighi and M. Uysal, "Survey on free space optical communication: a communication theory perspective," *IEEE Communication Surveys & Tutorials*, vol. 16, no. 4, pp. 2231-2258, 2014.

[38] W.-S. Tsai, H.-H. Lu, C.-Y. Li, T.-C. Lu, C.-H. Liao, C.-A. Chu, and P.-C. Peng, "A 20-m/40-Gb/s 1550-nm DFB LD-based FSO link," *IEEE Photonics Journal*, vol. 7, no. 6, Dec. 2015.

[39] D. L. Begley, "Free-space laser communications: A historical perspective," in Proc. of *Annual Meeting of the IEEE Lasers and Electro-Optics Society*, Nov. 2002, pp. 391-392.

[40] A. Malik and P. Singh, "Free space optics: current applications and future challenges," *International Journal of Optics*, vol. 2015, Sep. 2015.

[41] J. C. Juarez, A. Dwivedi, A. R. Hammons, S. D. Jones, V. Weerackody, and Ro. A. Nichols, "Free-space optical communications for next-generation military networks," *IEEE Communications Magazine*, vol. 44, no. 1, pp. 46-51, Nov. 2006.

[42] Y. Qin, T. T. Vu, and Y. Ban, "Toward an optimal algorithm for LiDAR waveform decomposition," *IEEE Geoscience and Remote Sensing Letters*, vol. 9, no. 3, pp. 482-486, May 2012.

[43] LIDAR. (2007) [Online]. Available: http://lidar.ihrc.fiu.edu/aboutlidar.html

[44] A. G. Sarigiannidis, M. Iloridou, P. Nicopolitidis, G. Papadimitriou, F.-N. Pavlidou, P. G. Sarigiannidis, M. D. Louta, and Vasileios Vitsas, "Architectures and bandwidth allocation schemes for hybrid wireless-optical networks," *IEEE Communication Surveys & Tutorials*, vol. 17, no. 1, pp. 427-468, Sep. 2014.

[45] A. Sevincer, A. Bhattarai, M. Bilgi, M. Yuksel, and Nezih Pala, "LIGHTNETs: smart LIGHTing and mobile optical wireless networks – a survey," *IEEE Communication Surveys & Tutorials*, vol. 15, no. 4, pp. 1620-1641, Apr. 2013.

[46] D. Karunatilaka, F. Zafar, V. Kalavally, and R. Parthiban, "LED based indoor visible light communications: state of the art," *IEEE Communication Surveys & Tutorials*, vol. 17, no. 3, pp. 1649-1678, March 2015.

[47] R.Q. Shaddad, A. B. Mohammad, S. A. Al-Gailani, A. M. Al-hetar, and M. A. Elmagzoub, "A survey on access technologies for broadband optical and wireless networks," *Journal of Network and Computer Applications*, vol. 41, pp. 459-4728, May 2014.

[48] S. Arnon, J. Barry, G. Karagiannidis, R. Schober, M. Uysal, *Advanced Optical wireless communication systems*, Cambridge University Press, 2012.

[49] A. M. Khalid, G. Cossu, R. Corsini, P. Choudhury, and E. Ciaramella, "1-Gb/s transmission over a phosphorescent white LED by using rate-adaptive discrete multitone modulation," *IEEE Photonics Journal*, vol. 4, no. 5, pp. 1465-1473, Oct. 2012.

[50] G. Cossu, R. Corsini, A. M. Khalid, P. Choudhury, and E. Ciaramella, "3.4 Gbit/s visible optical wireless transmission based on RGB LED," *Optics Express*, vol. 20, no. 26, pp. B501-B506, 2012.

[51] F. Zafar, M. Bakaul, and R. Parthiban, "Laser-diode-based visible light communication: toward gigabit class communication," *IEEE Communications Magazine*, vol. 55, no. 2, pp. 144-151, Feb. 2006.

[52] J. M. Kahn and J. R. Barry, "Wireless infrared communications," *Proceedings of the IEEE*, vol. 85, no. 2, pp. 265-298, February 1997.

[53] M. R. Feldman, S. C. Esener, C. C. Guest, and S. H. Lee, "Comparison between electrical and optical interconnect based on power and speed consideration," *Applied Optics*, vol. 27, no. 9, pp. 1742–1751, Sep. 1998.

[54] D. A. B. Miller, "Rationale and challenges for optical interconnects to electronic chips," *Proceedings of IEEE*, vol. 88, no. 6, pp. 728–749, Jun. 2000.

[55] C. Kachris and I. Tomkos, "A survey on optical interconnects for data centers," *IEEE Communication Surveys & Tutorials*, vol. 14, no. 4, pp. 1021–1036, Oct. 2012.

[56] M. A. Taubenblatt, "Optical interconnects for high-performance computing," *Journal of Lightwave Technology*, vol. 30, no. 4, pp. 448–457, Feb. 2012.

[57] F. Hanson and S. Radic, "High bandwidth underwater optical communication," *Applied Optics*, vol. 47, no. 2, pp. 277–283, Jan. 2008.

[58] C. Gabriel, M. A. Khalighi, S. Bourennane, P. Léon, and V. Rigaud, "Monte-Carlo-based channel characterization for underwater optical communication systems," *IEEE/OSA Journal of Optical Communications and Networking*, vol. 5, no. 1, pp. 1–12, Jan. 2013.

[59] F. R. Gfeller and U. Bapst, "Wireless in-house data communication via diffuse infrared radiation," *Proceedings of IEEE*, vol. 67, no. 11, pp. 1474–1486, Nov. 1979.

[60] M. K. Sichitiu and M. Kihl, "Inter-vehicle communication systems: A survey," *IEEE Communication Surveys & Tutorials*, vol. 10, no. 2, pp. 88–105, Jul. 2008.

[61] H. Hemmati, *Deep Space Optical Communications*. Hoboken, NJ, USA: Wiley-Interscience, 2006.

[62] V. W. S. Chan, "Optical satellite networks," *Journal of Lightwave Technology*, vol. 21, no. 11, pp. 2811–2827, Nov. 2003.

[63] "IEEE Standard for Local and Metropolitan Area Networks--Part 15.7: Short-range wireless optical communication using visible light," IEEE Std 802.15.7-2011, pp.1-309, Sep. 2011.

[64] ISO 21348 Definitions of Solar Irradiance Spectral Categories, International Organization for Standardization (ISO), 2017 [Online]. Available: http://www.spacewx.com/pdf/SET_21348_2004.pdf

[65] An Introduction to National Aeronautics and Space Administration Spectrum Management, NASA, 2017. [Online]. Available: https://www.nasa.gov/sites/default/files/atoms/files/spectrum_101.pdf

[66] D. E. Sunstein, "A Scatter communications link at ultraviolet frequencies," *Thesis*, MIT, Cambridge, MA, 1968.

[67] G. A. Shaw, Andrew M. Siegel, J. Model, and D. Greisokh, "Recent progress in short-range ultraviolet communication," in Proc. of *SPIE 5796, Unattended Ground Sensor Technologies and Applications*, vol. 5796, pp. 214-225, Jun. 2005.

[68] Energy Savings Forecast of Solid-State Lighting in General Illumination Applications. (2014) United States Department of Energy [Online]. Available: http://apps1.eere.energy.gov/buildings/publications/pdfs/ssl/energysavingsforecast14.pdf

[69] LASER vs. LED: What's the Difference? (2017) Acupuncture Technology News. [Online]. Available: https://www.miridiatech.com/news/2016/02/laser-vs-led-whats-the-difference/

[70] H. L. Minh, D. O'Brien, G. Faulkner, Olivier Bouchet, Mike Wolf, Liane Grobe, and Jianhui Li, "A 1.25-Gb/s indoor cellular optical wireless communications demonstrator," *IEEE Photonics Technology Letters*, vol. 22, no. 21, pp. 1598–1600, Nov. 2010.

[71] World Health Organization. (2017) [Online]. Available: http://www.who.int/uv/resources/fact/fs202laserpointers.pdf

[72] Optics 4 Engineers. (2017) [Online]. Available: http://www.optique-ingenieur.org/en/courses/OPI_ang_M01_C02/co/Contenu_04.html

[73] A. Neumann, J. J Wierer, W. Davis, Y. Ohno, S. R. J. Brueck, and J.Y. Tsao, "Four-color laser white illuminant demonstrating high color-rendering quality," *Optics Express*, vol. 19, no. S4, pp. A982-A990, July 2011.

[74] R. Boubezari, H. L. Minh, Z. Ghassemlooy, and A. Bouridane, "Smartphone camera based visible light communication," *Journal of Lightwave Technology*, vol. 34, no. 17, pp. 4120-4126, Sept. 2016.

[75] D. Karunatilaka, F. Zafar, V. Kalavally, and Rajendran Parthiban, "LED based indoor visible light communications: state of the art," *IEEE*

*Communications Surveys & Tutorials*, vol. 17, no. 3, pp. 1649-1678, 2015.

[76] T. Komine and M. Nakagawa, "Fundamental analysis for visible-light communication system using LED lightings," *IEEE Transactions on Consumer Electronics*, vol. 50, no. 1, pp. 100-107, Feb. 2004.

[77] D. C. O'Brien, L. Zeng, H. Le-Minh, G. Faulkner, J. W. Walewski, and S. Randel, "Visible light communications: challenges and possibilities," in Proc. of *International Symposium on Personal, Indoor and Mobile Radio Communications*, Cannes, 2008, pp. 1-5.

[78] H. Liu., H. Darabi, P. Banerjee, and J. Liu, "Survey of wireless indoor positioning techniques and systems," *IEEE Transactions on Systems, Man, and Cybernetics, Part C (Applications and Reviews)*, vol. 37, no. 6, pp. 1067-1080, Nov. 2007.

[79] L. U. Khan, "Visible light communication: applications, architecture, standardization and research challenges," *Digital Communications and Networks*, Aug. 2016.

[80] Visible Light Communications Consortium. (2010) [Online]. Available: http://www.vlcc.net/?ml_lang=en

[81] The IEEE Photonics Conference live stream. (2017) [Online]. Available: http://ieeetv.ieee.org/mobile/video/lifi-misconceptions-conceptions-and -opportunities-harald-haas-keynote-from-the-2016-ieee-photonics-conf erence

[82] M. Ayyash, H. Elgala, A. Khreishah, V. Jungnickel, T. Little, S. Shao, M. Rahaim, D. Schulz, J. Hilt, and Ro. Freund, "Coexistence of WiFi and LiFi toward 5G: concepts, opportunities, and challenges," *IEEE Communications Magazine*, vol. 54, no. 2, pp. 64-71, Feb. 2016.

[83] S. Li, A. Pandharipande, and F. M. J. Willems, "Unidirectional visible light communication and illumination with LEDs," *IEEE Sensors Journal*, vol. 16, no. 23, pp. 8617-8626, Dec. 2016.

[84] W. Yuanquan and C. Nan, "A high-speed bi-directional visible light communication system based on RGB-LED," *China Communications*, vol. 11, no. 9, pp. 40-44, March 2014.

[85] H. Haas, "High-speed wireless networking using visible light," *SPIE Newsroom*, April 19, 2013.

[86] Scientific Review, *Non-Ionizing Electromagnetic Radiation in the Radiofrequency Spectrum and its Effects on Human Health*, (2010). [Online]. Available: http://www.wireless-health.org.br/downloads/ LatinAmericanScienceReviewReport.pdf

[87] K. Xu, H.-Y. Yu, Y.-J. Zhu, and Y. Sun, "On the ergodic channel capacity for indoor visible light communication systems," *IEEE Access*, vol. 5, pp. 833-841, March 2017.

[88] Y. Yi, C. P. Li, and K. S. Lee, "Wavelength division-adaptive interference cancellation applied in OFDM visible light communication systems," *IETE Journal of Research*, vol. 58, no. 5, pp. 390-397, Sep. 2012.

[89] M. Z. Chowdhury, M. S. Uddin, and Y. M. Jang, "Dynamic channel allocation for class-based QoS provisioning and call admission in visible light communication," *The Arabian Journal for Science and Engineering*, vol. 39, no. 2, pp. 1007-1016, Feb. 2014.

[90] D. Yamanaka, S. Haruyama, and M. Nakagawa, "The design of high-speed image sensor chip for receiving the data of visible-light ID system," *IEICE Technical Report*, vol. 107, no. 300, pp. 97-102, Oct. 2007.

[91] T. Yamazato, M. Kinoshita, S. Arai, E. Souke, T. Yendo, T. Fujii, K. Kamakura, and H. Okada, "Vehicle motion and pixel illumination modeling for image sensor based visible light communication," *IEEE Journal on Selected Areas in Communications*, vol. 33, no. 9, pp. 1793-1805, Sep. 2015.

[92] M. Yoshino, S. Haruyama, and M. Nakagawa, "High-accuracy positioning system using visible LED lights and image sensor," in Proc. of *IEEE Radio and Wireless Symposium*, Jan. 2008, pp. 439-442.

[93] W. A. Cahyadi, Y. H. Kim, Y. Ho Chung, and C.-J. Ahn, "Mobile phone camera-based indoor visible light communications with rotation compensation," *IEEE Photonics Journal*, vol. 8, no. 2, pp. 1-8, April 2016.

[94] Y. Oike, M. Ikeda, and K. Asada, "A smart image sensor with high-speed feeble ID-beacon detection for augmented reality system," in Proc. of *European Solid-State Circuits Conference*, Sep. 2003, pp. 125-128.

[95] N. Matsushita, D. Hihara, T. Ushiro, S. Yoshimura, J. Rekimoto, and Y. Yamamoto, "ID CAM: A smart camera for scene capturing and ID recognition," in Proc. of *IEEE and ACM International Symposium on Mixed and Augmented Reality*, Oct. 2003, pp. 227-236.

[96] S. Itoh, I. Takai, M. S. Z. Sarker, M. Hamai, K. Yasutomi, M. Andoh, and S. Kawahito, "A CMOS image sensor for 10Mb/s 70m-range

[97] LED-based spatial optical communication," in Proc. of *IEEE International Solid-State Circuits Conference*, Feb. 2010, pp. 402-403.

[97] M. S. Z. Sarker, S. Itoh, M. Hamai, I. Takai, M. Andoh, K. Yasutomi, and S. Kawahito, "Design and implementation of a CMOS light pulse receiver cell array for spatial optical communications," *Sensors*, vol. 11, no. 2, pp. 2056-2076, Feb. 2011.

[98] S. Nishimoto, T. Nagura, T. Yamazato, T. Yendo, T. Fujii, H. Okada, and S. Arai, "Overlay coding for road-to-vehicle visible light communication using LED array and high-speed camera," in Proc. of *International IEEE Conference on Intelligent Transportation Systems*, Oct. 2011, pp. 1704-1709.

[99] T. Nguyen, A. Islam, and Y. M. Jang, "Region-of-interest signaling vehicular system using optical camera communications," *IEEE Photonics Journal*, vol. 9, no. 1, Feb. 2017.

[100] T. Nguyen, A. Islam, M. T. Hossan, and Y. M. Jang, "Current status and performance analysis of optical camera communication technologies for 5G networks," *IEEE Access*, vol. 5, no. 1, pp. 4574-4594, April 2017.

[101] B. Lin, Z. Ghassemlooy, C. Lin, X, Tang, Y. Li, and S. Zhang, "An indoor visible light positioning system based on optical camera communications," *IEEE Photonics Technology Letters*, vol. 29, no. 1, pp. 579-582, April 2017.

[102] S. Pergoloni, M. Biagi, S. Colonnese, R. Cusani, and G. Scarano, "A space-time RLS algorithm for adaptive equalization: the camera communication case," *Journal of Lightwave Technology*, vol. 35, no. 10, pp. 1811-1820, May 2017.

[103] T. Fujihashi, T. Koike-Akino, P. V. Orlik, and T. Watanabe, "Experimental throughput analysis in screen-camera visual MIMO communications," in Proc. of *IEEE Global Communications Conference*, Dec. 2016, pp. 1-7.

[104] I. Takai, T. Harada, M. Andoh, K. Yasutomi, K. Kagawa, and S. Kawahito, "Optical vehicle-to-vehicle communication system using LED transmitter and camera receiver," *IEEE Photonics Journal*, vol. 6, no. 5, pp. 1-14, Oct. 2014.

[105] I. Takai, S. Ito, K. Yasutomi, K. Kagawa, M. Andoh, and S. Kawahito, "LED and CMOS image sensor based optical wireless communication system for automotive applications," *IEEE Photonics Journal*, vol. 5, no. 5, Oct. 2013.

[106] J.-Y. Kim, S.-H. Yang, Y.-H. Son, and S.-K. Han, "High-resolution indoor positioning using light emitting diode visible light and camera image sensor," *IET Optoelectronics*, vol. 10. no. 5, pp. 184-192, 2016.

[107] J. Armstrong, Y. A. Sekercioglu, and A. Neild, "Visible light positioning: a roadmap for international standardization," *IEEE Communications Magazine*, vol. 51, no. 12, pp. 68-73, Dec. 2013.

[108] Canon: The Canon EOS-1Ds MARK II: the absolute pinnacle of D-SLR design and performance (2004).

[109] M. Schöberl, A. Brückner, S. Foessel, and A. Kaup, "Photometric limits for digital camera systems," *Journal of Electronic Imaging*, vol. 21, no. 2, pp. 020501-020503, Jun. 2012.

[110] M. A. Kashani, M. Uysal, and M. Kavehrad, "On the performance of MIMO FSO communications over double generalized gamma fading channels," in Proc. of *IEEE International Conference on Communications*, Jun. 2015, pp. 5144-5149.

[111] A. A. El-Malek, A. Salhab, S. Zummo, and M.-S. Alouini, "Effect of RF interference on the security-reliability trade-off analysis of multiuser mixed RF/FSO relay networks with power allocation," *Journal of Lightwave Technology*, vol. 35, no. 9, pp. 1490-1505, May 2017.

[112] H. Yuksel, S. Milner, and C. C. Davis, "Aperture averaging for optimizing receiver design and system performance on free-space optical communication links," *Journal of Optical Networking*, vol. 4, no. 8, pp. 462-475, Aug. 2005.

[113] R. Lange, B. Smutny, B. Wandernoth, R. Czichy, and D. Giggenbach, "142 km, 5.625 Gb/s free-space optical link based on homodyne BPSK modulation," in Proc. of *SPIE Free-Space Laser Communication Technologies*, March 2006, pp. 61050A.

[114] B. Smutny *et al.*, "5.6 Gbps optical inter-satellite communication link," in Proc. of *SPIE Free-Space Laser Communication Technologies*, Feb. 2009.

[115] D. Killinger, "Free space optics for laser communication through the air," *Optics and Photonics News*, vol. 13, no. 10, pp. 36-42, Oct. 2002.

[116] H. Henniger and O. Wilfert, "An introduction to free-space optical communications," *Radio Engineering*, vol. 19, no. 2, pp. 203-212, Jun. 2010.

[117] X. Liu, "Free-space optics optimization models for building sway and atmospheric interference using variable wavelength," *IEEE*

*Transactions on Communications*, vol. 57, no. 2, pp. 492-498, Feb., 2009.

[118] M.-C. Jeung *et al.*, "8 × 10-Gb/s terrestrial optical free-space transmission over 3.4 km using an optical repeater," *IEEE Photonics Technology Letters*, vol. 15, no. 1, pp. 171-173, Jan. 2003.

[119] E. Ciaramella *et al.*, "1.28-Tb/s (32 × 40 Gb/s) free-space optical WDM transmission system," *IEEE Photonics Technology Letters*, vol. 21, no. 16, pp. 1121-1123, Aug. 2009.

[120] Y. Kaymak, R. Rojas-Cessa, J. Feng, N. Ansari, and M. C. Zhou, "On divergence-angle efficiency of a laser beam in free-space optical communications for high-speed trains," *IEEE Transactions on Vehicular Technology*, vol. PP, no. 99, pp. 1-1, March 2017.

[121] A. K. Majumdar and J. C. Ricklin, *Free-Space Laser Communications: Principles and Advances*. New York, NY, USA: Springer-Verlag, Dec. 2010.

[122] S. S. Muhammad, B. Flecker, E. Leitgeb, and M. Gebhart, "Characterization of fog attenuation in terrestrial free space optical links," *Optical Engineering*, vol. 46, no. 6, pp. 066001-066010, Jun. 2007.

[123] M. L. B. Riediger, R. Schober, and L. Lampe, "Fast multiple-symbol detection for free-space optical communications," *IEEE Transactions on Communications*, vol. 57, no. 4, pp. 1119-1128, April 2009.

[124] D. Rollins *et al.*, "Background light environment for free-space optical terrestrial communications links," in Proc. of *SPIE, Optical Wireless Communication*. Dec. 2002, pp. 99-110.

[125] H. Kaushal and G. Kaddoum, "Optical communication in space: challenges and mitigation techniques," *IEEE Communications Surveys & Tutorials*, vol. 19, no. 1, pp. 57-97, 2017.

[126] S. Zhang *et al.*, "1.5 Gbit/s multi-channel visible light communications using CMOS-controlled GaN- ased LEDs," *Journal of Lightwave Technology*, vol. 31, no. 8, pp. 1211-1216, Aug. 2013.

[127] A. S. Ghazy, H. A. I. Selmy, and H. M. H. Shalaby, "Fair resource allocation schemes for cooperative dynamic free-space optical networks," *IEEE/OSA Journal of Optical Communications and Networking*, vol. 8, no. 11, pp. 822-834, Nov. 2016.

[128] M. P. Peloso and I. Gerhardt, "Statistical tests of randomness on quantum keys distributed through a free-space channel coupled to daylight noise," vol. 31, no. 23, pp. 3794-3805, Dec. 2013.

[129] W. J. Miniscalco and S. A. Lane, "Optical space-time division multiple access," *Journal of Lightwave Technology*, vol. 30, no. 11, pp. 1771-1785, Nov. 2012.

[130] K. Su, L. Moeller, R. B. Barat, and J. F. Federici, "Experimental comparison of performance degradation from terahertz and infrared wireless links in fog," *Journal of the Optical Society of America A*, vol. 29, no. 2, pp. 179-184, Feb. 2012.

[131] H. A. Willebrand and B. S. Ghuman, "Fiber optics without fiber," *IEEE Spectrum*, vol. 40, no. 8, pp. 41-45, Aug. 2001.

[132] E. Leitgeb *et al.*, "Current optical technologies for wireless access," in Proc. of *International Conference on Telecommunications*, Jun. 2009, pp. 7-17.

[133] D. Kedar and S. Arnon, "Urban optical wireless communication networks: The main challenges and possible solutions," *IEEE Communications Magazine*, vol. 42, no. 5, pp. 2-7, May 2004.

[134] M. Weib, M. Huchard, A. Stohr, B. Charbonnier, S. Fedderwitz, and D. S. Jager, "60-GHz Photonic Millimeter-Wave Link for Short-to Medium-Range Wireless Transmission Up to 12.5 Gb/s," *Journal of Lightwave Technology*, vol. 26, no. 15, pp. 2424-2429, Aug. 2008.

[135] Autonomos Labs: Autonomous Cars from Berlin. (2015) [Online]. Available: http://autonomos-labs.com/research/

[136] A. P. Cracknell, *Introduction to Remote Sensing*, CRC Press, London: Taylor and Francis.

[137] Lidar 101: An Introduction to Lidar Technology, Data, and Applications. (Nov. 2012) NOAA Coastal Services Center. [Online]. Available: https://coast.noaa.gov/data/digitalcoast/pdf/lidar-101.pdf

[138] K. Zhang and F. Christopher, "Road grade estimation for on-road vehicle emissions modeling using light detection and ranging data," *Journal of the Air & Waste Management Association*, vol. 56, no. 6, pp. 777-788, 2012.

[139] The Royal Institution of Australia. (2017) Ri Aus. [Online]. Available https://riaus.org.au/

[140] LiDAR sensor provider. (2017) Velodyne LiDAR. [Online]. Available: http://www.velodynelidar.com/

[141] G. Gigli and N. Casagli, "Semi-automatic extraction of rock mass structural data from high resolution LIDAR point clouds," *International Journal of Rock Mechanics and Mining Sciences,* vol. 48, pp. 187-198, 2011.

[142] LiDAR solutions. IfTAS GmbH (2015). [Online]. Available: http://www.iftas.de/Main/Solutions.

[143] V. Sharma and Sushank, "High speed CO-OFDM-FSO transmission system," *International Journal for Light and Electron Optics,* vol. 125, no. 6, pp. 1761-1763, March 2014.

[144] X. Song, F. Yang, and J. Cheng, "Subcarrier BPSK modulated FSO communications with pointing errors," in Proc. of *IEEE Wireless Communications and Networking Conference*, April 2013, pp. 4261-4265.

[145] S.-H. Chen and C.-W. Chow, "Color-shift keying and code-division multiple-access transmission for RGB-LED visible light communications using mobile phone camera," *IEEE Photonics Journal*, vol. 6, no. 6, pp. 1-6, Dec. 2014.

[146] Radio-electronics website. (2017) [Online]. Available: http://www.radio-electronics.com/info/wireless/wi-fi/ieee-802-11ac-gig abit.php

[147] G. Pan, J. Ye, and Z. Ding, "Secure hybrid VLC-RF systems with light energy harvesting," *IEEE Transactions on Communications*, vol. 65, no. 10, pp. 4348-4359, Oct. 2017.

[148] C. Chen, S. Videv, D. Tsonev, and Harald Haas, "Fractional frequency reuse in DCO-OFDM-based optical attocell networks," *Journal of Lightwave Technology*, vol. 33, no. 19, pp. 3986-4000, Oct. 1, 2015.

[149] Z. Zeng, S. Fu, H. Zhang, Y. Dong, and J. Cheng, "A survey of underwater optical wireless communications," *IEEE Communication Surveys & Tutorials*, vol. 19, no. 1, pp. 204-238, 2017.

[150] H. Furuhata, "Electromagnetic interferences of electric medical equipment from hand-held radio communication equipment," in Proceedings of *International Symposium on Electromagnetic Compatibility*, May 1999, pp. 468–71.

[151] Y. L. Morgan, "Notes on DSRC & WAVE standards suite: its architecture, design, and characteristics," *IEEE Communications Surveys & Tutorials*, vol. 12, no. 4, pp. 504-518, 2010.

[152] T. Yamazato, N. Kawagita, H. Okada, T. Fujii, T. Yendo, S. Arai, and K. Kamakura, "The uplink visible light communication beacon system for universal traffic management," *IEEE Access*, vol. 5, pp. 22282 – 22290, Nov. 2017.

[153] G. Chmaj and H. Selvaraj, "Distributed processing applications for UAV/drones: a survey," *Progress in Systems Engineering*, vol. 366, pp. 449–454, 2015.